\documentclass[sn-mathphys-num,iicol]{sn-jnl}

\usepackage[utf8]{inputenc}
\usepackage[english]{babel}
\usepackage{amsmath}
\usepackage{subfigure}
\usepackage{url}
\usepackage{units}
\usepackage{booktabs}
\usepackage{makecell}
\usepackage{multirow}
\usepackage[table]{xcolor}
\usepackage{float}

\usepackage[title]{appendix}%

\newcommand{\cawo}[1][ ]{CaWO$_4$#1}
\newcommand{\alo}[1][ ]{Al$_2$O$_3$#1}
\newcommand{\cevns}[1][ ]{CE$\nu$NS#1}

\let\orcidlogo\undefined
\usepackage{orcidlink}
\usepackage{xparse}

\begin{document}

\title[Characterization of the Low Energy Excess using a NUCLEUS \alo  detector]{\textbf{Characterization of the Low Energy Excess using a NUCLEUS \alo  detector}}




%
\newcount\authorStyle

\authorStyle=1

%
\newcount\instByName

\instByName=0

\newcommand{\iNUCLEUScontactEmail}{Contact E-Mail of NUCLEUS Collaboration : info@nucleus-experiment.org}

\ifcase\authorStyle
    \collaboration{NUCLEUS Collaboration}
\else
\fi

\ifnum\authorStyle=1
\else
    \newcommand{\orgdiv}[1]{#1}%
    \newcommand{\orgname}[1]{#1}%
    \newcommand{\orgaddress}[1]{#1}%
    \newcommand{\street}[1]{#1}%
    \newcommand{\postcode}[1]{#1}%
    \newcommand{\city}[1]{#1}%
    \newcommand{\state}[1]{#1}%
    \newcommand{\country}[1]{#1}%
\fi





\newcommand{\iMBI}{%
    \orgname{Marietta-Blau-Institut f{\"u}r Teilchenphysik der {\"O}sterreichischen Akademie der Wissenschaften}, 
    \orgaddress{
        \street{Dominikanerbastei~16}, 
        \city{Wien}, 
        \postcode{A-1010}, 
        \country{Austria}%
        }%
    }

\newcommand{\iTUW}{%
    \orgdiv{Atominstitut}, 
    \orgname{Technische Universit\"at Wien}, 
    \orgaddress{
        \street{Stadionallee~2}, 
        \city{Wien}, 
        \postcode{A-1020}, 
        \country{Austria}%
        }%
    }


\newcommand{\iCEA}{%
    \orgdiv{IRFU}, 
    \orgname{CEA, Universit\'{e} Paris-Saclay}, 
    \orgaddress{
        \street{B\^{a}timent 141}, 
        \city{Gif-sur-Yvette}, 
        \postcode{F-91191}, 
        \country{France}%
        }%
    }

\newcommand{\iEdF}{%
    \orgdiv{Centre nucl{\'e}aire de production d'{\'}electricit{\'e} de Chooz, Service Automatismes-Essais}, 
    \orgname{{\'E}lectricit{\'e} de France}, 
    \orgaddress{
        \street{}, 
        \city{Givet}, 
        \postcode{F-08600}, 
        \country{France}%
        }%
    }


\newcommand{\iMPIK}{%
    \orgname{Max-Planck-Institut für Kernphysik}, 
    \orgaddress{%
        \street{Saupfercheckweg 1}, 
        \city{Heidelberg}, 
        \postcode{D-69117}, 
        \country{Germany}%
        }%
    }

\newcommand{\iMPP}{%
    \orgname{Max-Planck-Institut f{\"u}r Physik}, 
    \orgaddress{
        \street{Boltzmannstra{\ss}e~8}, 
        \city{Garching}, 
        \postcode{D-85748}, 
        \country{Germany}%
        }%
    }

\newcommand{\iTUM}{%
    \orgdiv{Physik-Department, TUM School of Natural Sciences}, 
    \orgname{Technische Universit\"at M\"unchen}, 
    \orgaddress{
        \street{James-Franck-Straße 1}, 
        \city{Garching}, 
        \postcode{D-85748}, 
        \country{Germany}%
        }%
    }


\newcommand{\iINFNRoma}{%
    \orgname{Istituto Nazionale di Fisica Nucleare -- Sezione di Roma}, 
    \orgaddress{
        \street{Piazzale Aldo Moro 2}, 
        \city{Roma}, 
        \postcode{I-00185}, 
        \country{Italy}%
        }%
    }

\newcommand{\iSapienza}{%
    \orgdiv{Dipartimento di Fisica}, 
    \orgname{Sapienza Universit\`{a} di Roma}, 
    \orgaddress{
        \street{Piazzale Aldo Moro 5}, 
        \city{Roma}, 
        \postcode{I-00185}, 
        \country{Italy}%
        }%
    }
    
\newcommand{\iINFNTorVergata}{%
    \orgname{Istituto Nazionale di Fisica Nucleare -- Sezione di Roma "Tor Vergata"}, 
    \orgaddress{
        \street{Via della Ricerca Scientifica 1}, 
        \city{Roma}, 
        \postcode{I-00133}, 
        \country{Italy}%
        }%
    }

\newcommand{\iTorVergata}{%
    \orgdiv{Dipartimento di Fisica}, 
    \orgname{Universit\`{a} di Roma "Tor Vergata"}, 
    \orgaddress{
        \street{Via della Ricerca Scientifica 1}, 
        \city{Roma}, 
        \postcode{I-00133}, 
        \country{Italy}%
        }%
    }

\newcommand{\iCNR}{%
    \orgdiv{Istituto di Nanotecnologia}, 
    \orgname{Consiglio Nazionale delle Ricerche}, 
    \orgaddress{
        \street{Piazzale Aldo Moro 5}, 
        \city{Roma}, 
        \postcode{I-00185}, 
        \country{Italy}%
        }%
    }

\newcommand{\iINFNFerrara}{%
    \orgname{Istituto Nazionale di Fisica Nucleare -- Sezione di Ferrara}, 
    \orgaddress{
        \street{Via Giuseppe Saragat 1c}, 
        \city{Ferrara}, 
        \postcode{I-44122}, 
        \country{Italy}%
        }%
    }

\newcommand{\iFerrara}{%
    \orgdiv{Dipartimento di Fisica}, 
    \orgname{Universit{\`a} di Ferrara}, 
    \orgaddress{
        \street{Via Giuseppe Saragat 1}, 
        \city{Ferrara}, 
        \postcode{I-44122}, 
        \country{Italy}%
        }%
    }

\newcommand{\iINFNLnGS}{%
    \orgname{Istituto Nazionale di Fisica Nucleare -- Laboratori Nazionali del Gran Sasso}, 
    \orgaddress{
        \street{Via Giovanni Acitelli 22}, 
        \city{Assergi (L’Aquila)}, 
        \postcode{I-67100}, 
        \country{Italy}%
        }%
    }

\newcommand{\iBicocca}{%
    \orgdiv{Dipartimento di Fisica}, 
    \orgname{Universit\`{a} di Milano Bicocca}, 
    \orgaddress{
        \street{}, 
        \city{Milan}, 
        \postcode{I-20126}, 
        \country{Italy}%
        }%
    }


\newcommand{\iCoimbra}{%
    \orgdiv{LIBPhys-UC, Departamento de Fisica}, 
    \orgname{Universidade de Coimbra}, 
    \orgaddress{
        \street{Rua Larga 3004-516}, 
        \city{Coimbra}, 
        \postcode{P3004-516}, 
        \country{Portugal}%
        }%
    }

%
%
\newcommand{\iAlsoAtCoimbra}{Also at \iCoimbra}
\newcommand{\iNowAtMPIK}{Now at \iMPIK}
\newcommand{\iNowAtLNGS}{Now at \iINFNLnGS}
\newcommand{\iNowAtKiutra}{Now at kiutra GmbH, Fl{\"o}{\ss}ergasse 2, D-81369 Munich, Germany}


\ifnum\instByName=1
    
    \newcommand{\MBI}{MBI}
    \newcommand{\TUW}{TUW}

    \newcommand{\CEA}{CEA}
    \newcommand{\EdF}{EdF}

    \newcommand{\MPIK}{MPIK}
    \newcommand{\MPP}{MPP}
    \newcommand{\TUM}{TUM}

    \newcommand{\INFNRoma}{INFNRoma}
    \newcommand{\Sapienza}{Sapienza}
    \newcommand{\INFNTorVergata}{INFNTorVergata}
    \newcommand{\TorVergata}{TorVergata}
    \newcommand{\CNR}{CNR}
    \newcommand{\INFNFerrara}{INFNFerrara}
    \newcommand{\Ferrara}{Ferrara}
    \newcommand{\INFNLnGS}{INFNLnGS}
    \newcommand{\Bicocca}{Bicocca}    

    \newcommand{\Coimbra}{Coimbra}
\else    
    
    \newcommand{\MBI}{3}
    \newcommand{\TUW}{1}

    \newcommand{\CEA}{5}
    \newcommand{\EdF}{ERROR}

    \newcommand{\MPIK}{ERROR}
    \newcommand{\MPP}{2}
    \newcommand{\TUM}{8}

    \newcommand{\INFNRoma}{6}
    \newcommand{\Sapienza}{7}
    
    \newcommand{\INFNTorVergata}{4}
    \newcommand{\TorVergata}{9}
    
    \newcommand{\CNR}{ERROR}
    
    \newcommand{\Ferrara}{10}
    \newcommand{\INFNFerrara}{11}
    
    \newcommand{\INFNLnGS}{ERROR}
    
    \newcommand{\Bicocca}{ERROR}    

    \newcommand{\Coimbra}{ERROR}

\fi




\ifcase\authorStyle
    \author{H.~Abele~\orcidlink{0000-0002-6832-9051}}
    \affiliation{\iTUW}
\or
    \author[\TUW]{\fnm{H.}~
    \sur{Abele}~
    \orcidlink{0000-0002-6832-9051}
        }
\else
\fi

\ifcase\authorStyle
    \author{G.~Angloher}
     \affiliation{\iMPP}
\or
    \author[\MPP]{
    \fnm{G.}~
    \sur{Angloher} 
    }
\else
\fi

\ifcase\authorStyle
    \author{B.~Arnold}
    \affiliation{\iMBI}
\or
    \author[\MBI]{
        \fnm{B.}~
        \sur{Arnold}
        }
\else
\fi
        
\ifcase\authorStyle
    \author{M.~Atzori~Corona~\orcidlink{0000-0001-5092-3602}}
    \affiliation{\iINFNTorVergata}
\or
    \author[\INFNTorVergata]{
    \fnm{M.}~
    \sur{Atzori~Corona}~
    \orcidlink{0000-0001-5092-3602}
    }
\else
\fi

\ifcase\authorStyle
    \author{A.~Bento~\orcidlink{0000-0002-3817-6015}}
    \thanks{\iAlsoAtCoimbra}
     \affiliation{\iMPP}
\or
    \author[\MPP]{
        \fnm{A.}~
        \sur{Bento}~
        \orcidlink{0000-0002-3817-6015}~
        \textsuperscript{\symAlsoAtCoimbra,\,}%
        }
    \else
\fi

\ifcase\authorStyle
    \author{E.~Bossio~\orcidlink{0000-0001-9304-1829}}
    \affiliation{\iCEA}
\or
    \author[\CEA]{
        \fnm{E.}~
        \sur{Bossio}~
        \orcidlink{0000-0001-9304-1829}
        }
    \else
\fi
    
\ifcase\authorStyle
    \author{F.~Buchsteiner}
    \affiliation{\iMBI}
\or
    \author[\MBI]{
        \fnm{F.}~
        \sur{Buchsteiner}
        }
    \else
\fi
    
\ifcase\authorStyle
    \author{J.~Burkhart~\orcidlink{0000-0002-1989-7845}}
    \affiliation{\iMBI}
\or
    \author[\MBI]{
        \fnm{J.}~
        \sur{Burkhart}~
        \orcidlink{0000-0002-1989-7845}
        }
\else
\fi
    

\ifcase\authorStyle
    \author{F.~Cappella~\orcidlink{0000-0003-0900-6794}}
    \affiliation{\iINFNRoma}
\or
    \author[\INFNRoma]{
        \fnm{F.}~
        \sur{Cappella}~
        \orcidlink{0000-0003-0900-6794}
        }
\else
\fi
    
\ifcase\authorStyle
    \author{M.~Cappelli~\orcidlink{0009-0002-6148-5964}}
    \affiliation{\iSapienza}
    \affiliation{\iINFNRoma}
\or
    \author[\Sapienza, \INFNRoma]{
        \fnm{M.}~
        \sur{Cappelli}~
        \orcidlink{0009-0002-6148-5964}
        }
\else
\fi
    

\ifcase\authorStyle
    \author{N.~Casali~\orcidlink{0000-0003-3669-8247}}
    \affiliation{\iINFNRoma}
\or
    \author[\INFNRoma]{
        \fnm{N.}~
        \sur{Casali}~
        \orcidlink{0000-0003-3669-8247}
        }
\else
\fi
    
\ifcase\authorStyle
    \author{R.~Cerulli~\orcidlink{0000-0003-2051-3471}}
    \affiliation{\iINFNTorVergata}
\or
    \author[\INFNTorVergata]{
        \fnm{R.}~
        \sur{Cerulli}~
        \orcidlink{0000-0003-2051-3471}
        }
\else
\fi
    

\ifcase\authorStyle
    \author{A.~Cruciani~\orcidlink{0000-0003-2247-8067}}
    \affiliation{\iINFNRoma}
\or
    \author[\INFNRoma]{
        \fnm{A.}~
        \sur{Cruciani}~
        \orcidlink{0000-0003-2247-8067}
        }
\else
\fi
    
\ifcase\authorStyle
    \author{G.~Del~Castello~\orcidlink{0000-0001-7182-358X}}
    \thanks{Corresponding author: G.~Del~Castello}
    \affiliation{\iINFNRoma}
\or
    \author*[\INFNRoma]{
        \fnm{G.}~
        \sur{Del~Castello}~
        \email{Del~Castello, G.: giorgio.delcastello@roma1.infn.it}~
        \orcidlink{0000-0001-7182-358X}
        }
\else
\fi

\ifcase\authorStyle
    \author{M.~del~Gallo~Roccagiovine~\orcidlink{0009-0006-5861-7443}}
    \affiliation{\iSapienza}
    \affiliation{\iINFNRoma}
\or
    \author[\Sapienza, \INFNRoma]{
        \fnm{M.}~
        \sur{del~Gallo~Roccagiovine}~
        \orcidlink{0009-0006-5861-7443}
        }
\else
\fi
    

\ifcase\authorStyle
    \author{S.~Dorer~\orcidlink{0009-0001-1670-5780}}
    \affiliation{\iTUW}
\or
    \author[\TUW]{
        \fnm{S.}~
        \sur{Dorer}~
        \orcidlink{0009-0001-1670-5780}
        }
\else
\fi
    
\ifcase\authorStyle
    \author{A.~Erhart~\orcidlink{0000-0002-8721-177X}}
    \affiliation{\iTUM}
\or
    \author[\TUM]{
        \fnm{A.}~
        \sur{Erhart}~
        \orcidlink{0000-0002-8721-177X}
        }
\else
\fi
    
\ifcase\authorStyle
    \author{M.~Friedl~\orcidlink{0000-0002-7420-2559}}
    \affiliation{\iMBI}
\or
    \author[\MBI]{
        \fnm{M.}~
        \sur{Friedl}~
        \orcidlink{0000-0002-7420-2559}}
\else
\fi

\ifcase\authorStyle
    \author{S.~Fichtinger}
    \affiliation{\iMBI}
\or
    \author[\MBI]{
        \fnm{S.}~
        \sur{Fichtinger}
        }
\else
\fi


\ifcase\authorStyle
    \author{V.M.~Ghete~\orcidlink{0000-0002-9595-6560}}
    \affiliation{\iMBI}
\or
    \author[\MBI]{
        \fnm{V.M.}~
        \sur{Ghete}~
        \orcidlink{0000-0002-9595-6560}
        }
\else
\fi

\ifcase\authorStyle
    \author{M.~Giammei~\orcidlink{0009-0006-9104-2055}}
    \affiliation{\iTorVergata}
    \affiliation{\iINFNTorVergata}
\or
    \author[\TorVergata, \INFNTorVergata]{
        \fnm{M.}~
        \sur{Giammei}~
        \orcidlink{0009-0006-9104-2055}
        }
\else
\fi

\ifcase\authorStyle
    \author{C.~Goupy~\orcidlink{0000-0003-4954-5311}}
    \thanks{\iNowAtMPIK}
    \affiliation{\iCEA}
\or
    \author[\CEA]{
        \fnm{C.}~
        \sur{Goupy}~
        \orcidlink{0000-0003-4954-5311}~
        \textsuperscript{\symNowAtMPIK,\,}%
        }
\else
\fi


\ifcase\authorStyle
    \author{J.~Hakenm{\"u}ller~\orcidlink{0000-0003-0470-3320}}
    \affiliation{\iMBI}
\or
    \author[\MBI]{
        \fnm{J.}~
        \sur{Hakenm{\"u}ller}~
        \orcidlink{0000-0003-0470-3320}}
\else
\fi

\ifcase\authorStyle
    \author{D.~Hauff}
     \affiliation{\iMPP}
    \affiliation{\iTUM}
\or
    \author[\MPP, \TUM]{
        \fnm{D.}~
        \sur{Hauff}
        }
\else
\fi
    
\ifcase\authorStyle
    \author{F.~Jeanneau~\orcidlink{0000-0002-6360-6136}}
    \affiliation{\iCEA}
\or
    \author[\CEA]{
        \fnm{F.}~
        \sur{Jeanneau}~
        \orcidlink{0000-0002-6360-6136}}
\else
\fi

\ifcase\authorStyle
    \author{E.~Jericha~\orcidlink{0000-0002-8663-0526}}
    \affiliation{\iTUW}
\or
    \author[\TUW]{
        \fnm{E.}~
        \sur{Jericha}
        }
\else
\fi

\ifcase\authorStyle
    \author{M.~Kaznacheeva~\orcidlink{0000-0002-2712-1326}}
    \thanks{Corresponding author: M.~Kaznacheeva}
    \affiliation{\iTUM}
\or
    \author*[\TUM]{
        \fnm{M.}~
        \sur{Kaznacheeva}~
        \email{\newline Kaznacheeva, M.: margarita.kaznacheeva@tum.de}~
        \orcidlink{0000-0002-2712-1326}
        }
\else
\fi


\ifcase\authorStyle
    \author{H.~Kluck~\orcidlink{0000-0003-3061-3732}}
    \affiliation{\iMBI}
\or
\author[\MBI]{
    \fnm{H.}~
    \sur{Kluck}~
    \orcidlink{0000-0003-3061-3732}
    }
\else
\fi

\ifcase\authorStyle
    \author{A.~Langenk{\"a}mper}
     \affiliation{\iMPP}
\or
    \author[\MPP]{
        \fnm{A.}~
        \sur{Langenk\"{a}mper} 
        }
\else
\fi

\ifcase\authorStyle
    \author{T.~Lasserre~\orcidlink{0000-0002-4975-2321}}
    \thanks{\iNowAtMPIK}
    \affiliation{\iCEA}
    \affiliation{\iTUM}
\or
    \author[\CEA, \TUM]{
        \fnm{T.}~
        \sur{Lasserre}~
        \orcidlink{0000-0002-4975-2321}~
        \textsuperscript{\symNowAtMPIK,\,}%
        }
\else
\fi

\ifcase\authorStyle
    \author{D.~Lhuillier~\orcidlink{0000-0003-2324-0149}}
    \affiliation{\iCEA}
\or
    \author[\CEA]{
        \fnm{D.}~
        \sur{Lhuillier}~
        \orcidlink{0000-0003-2324-0149}
        }
\else
\fi

\ifcase\authorStyle
    \author{M.~Mancuso~\orcidlink{0000-0001-9805-475X}}
     \affiliation{\iMPP}
\or
    \author[\MPP]{
        \fnm{M.}~
        \sur{Mancuso}~
        \orcidlink{0000-0001-9805-475X}
        }
\else
\fi

\ifcase\authorStyle
    \author{R.~Martin}
    \affiliation{\iCEA}
    \affiliation{\iTUW}
\or
    \author[\CEA, \TUW]{
        \fnm{R.}~
        \sur{Martin}
        }
\else
\fi

\ifcase\authorStyle
    \author{B.~Mauri}
     \affiliation{\iMPP}
\or
    \author[\MPP]{
        \fnm{B.}~
        \sur{Mauri} 
        }
\else
\fi

\ifcase\authorStyle
    \author{A.~Mazzolari}
    \affiliation{\iFerrara}
    \affiliation{\iINFNFerrara}
\or
    \author[\Ferrara, \INFNFerrara]{
        \fnm{A.}~
        \sur{Mazzolari} 
        }
\else
\fi

\ifcase\authorStyle
    \author{L.~McCallin}
    \affiliation{\iCEA}
\or
    \author[\CEA]{
        \fnm{L.}~
        \sur{McCallin} 
        }
\else
\fi


\ifcase\authorStyle
    \author{H.~Neyrial}
    \affiliation{\iCEA}
\or
    \author[\CEA]{
        \fnm{H.}~
        \sur{Neyrial} 
        }
\else
\fi

\ifcase\authorStyle
    \author{C.~Nones}
    \affiliation{\iCEA}
\or
    \author[\CEA]{
        \fnm{C.}~
        \sur{Nones} 
        }
\else
\fi

\ifcase\authorStyle
    \author{L.~Oberauer}
    \affiliation{\iTUM}
\or
    \author[\TUM]{
        \fnm{L.}~
        \sur{Oberauer} 
        }
\else
\fi



\ifcase\authorStyle
    \author{L.~Peters~\orcidlink{0000-0002-1649-8582}}
    \thanks{\iNowAtMPIK}
    \affiliation{\iTUM}
    \affiliation{\iCEA}
\or
    \author[\TUM, \CEA]{
        \fnm{L.}~
        \sur{Peters}~
        \orcidlink{0000-0002-1649-8582}~
        \textsuperscript{\symNowAtMPIK,\,}%
        }
\else
\fi

\ifcase\authorStyle
    \author{F.~Petricca~\orcidlink{0000-0002-6355-2545}}
    \affiliation{\iMPP}
\or
    \author[\MPP]{
        \fnm{F.}~
        \sur{Petricca}~
        \orcidlink{0000-0002-6355-2545}
        }
\else
\fi

\ifcase\authorStyle
    \author{W.~Potzel}
    \affiliation{\iTUM}
\or
    \author[\TUM]{
        \fnm{W.}~
        \sur{Potzel} 
        }
\else
\fi

\ifcase\authorStyle
    \author{F.~Pr\"{o}bst}
     \affiliation{\iMPP}
\or
    \author[\MPP]{
        \fnm{F.}~
        \sur{Pr\"{o}bst} 
        }
\else
\fi

\ifcase\authorStyle
    \author{F.~Pucci~\orcidlink{0000-0003-3782-2393}}
    \thanks{\iINFNLnGS}
     \affiliation{\iMPP}
\or
    \author[\MPP]{
        \fnm{F.}~
        \sur{Pucci}~
        \orcidlink{0000-0003-3782-2393}~
        \textsuperscript{\symNowAtLNGS,\,}%
        }
\else
\fi

\ifcase\authorStyle
    \author{F.~Reindl~\orcidlink{0000-0003-0151-2174}}
    \affiliation{\iMBI}
    \affiliation{\iTUW}
\or
    \author[\MBI, \TUW]{
        \fnm{F.}~
        \sur{Reindl}~
        \orcidlink{0000-0003-0151-2174}
        }
\else
\fi


\ifcase\authorStyle
    \author{M.~Romagnoni}
    \affiliation{\iFerrara}
    \affiliation{\iINFNFerrara}
\or
    \author[\Ferrara, \INFNFerrara]{
        \fnm{M.}~
        \sur{Romagnoni} 
        }
\else
\fi

\ifcase\authorStyle
    \author{J.~Rothe~\orcidlink{0000-0001-5748-7428}}
    \thanks{\iNowAtKiutra}
    \affiliation{\iTUM}
\or
    \author[\TUM]{
        \fnm{J.}~
        \sur{Rothe}~
        \orcidlink{0000-0001-5748-7428}~
        \textsuperscript{\symNowAtKiutra,\,}%
        }
\else
\fi

\ifcase\authorStyle
    \author{N.~Schermer~\orcidlink{0009-0004-4213-5154}}
    \affiliation{\iTUM}
\or
    \author[\TUM]{
        \fnm{N.}~
        \sur{Schermer}~
        \orcidlink{0009-0004-4213-5154}
        }
\else
\fi

\ifcase\authorStyle
    \author{J.~Schieck~\orcidlink{0000-0002-1058-8093}}
    \affiliation{\iMBI}
    \affiliation{\iTUW}
\or
    \author[\MBI, \TUW]{
        \fnm{J.}~
        \sur{Schieck}~
        \orcidlink{0000-0002-1058-8093}
        }
\else
\fi

\ifcase\authorStyle
    \author{S.~Sch\"{o}nert~\orcidlink{0000-0001-5276-2881}}
    \affiliation{\iTUM}
\or
    \author[\TUM]{
        \fnm{S.}~
        \sur{Sch\"{o}nert}~
        \orcidlink{0000-0001-5276-2881}
        }
\else
\fi

\ifcase\authorStyle
    \author{C.~Schwertner}
    \affiliation{\iMBI}
    \affiliation{\iTUW}
\or
    \author[\MBI, \TUW]{
        \fnm{C.}~
        \sur{Schwertner} 
        }
\else
\fi

\ifcase\authorStyle
    \author{L.~Scola}
    \affiliation{\iCEA}
\or
    \author[\CEA]{
        \fnm{L.}~
        \sur{Scola} 
        }
\else
\fi

\ifcase\authorStyle
    \author{G.~Soum-Sidikov~\orcidlink{0000-0003-1900-1794}}
    \affiliation{\iCEA}
\or
    \author[\CEA]{
        \fnm{G.}~
        \sur{Soum-Sidikov}~
        \orcidlink{0000-0003-1900-1794}
        }
\else
\fi

\ifcase\authorStyle
    \author{L.~Stodolsky}
     \affiliation{\iMPP}
\or
    \author[\MPP]{
        \fnm{L.}~
        \sur{Stodolsky} 
        }
\else
\fi

\ifcase\authorStyle
    \author{A.~Schr{\"o}der~\orcidlink{0009-0005-1598-1635}}
    \affiliation{\iTUM}
\or
    \author[\TUM]{
        \fnm{A.}~
        \sur{Schr{\"o}der}~
        \orcidlink{0009-0005-1598-1635}
        }
\else
\fi

\ifcase\authorStyle
    \author{R.~Strauss~\orcidlink{0000-0002-5589-9952}}
    \affiliation{\iTUM}
\or
    \author[\TUM]{
        \fnm{R.}~
        \sur{Strauss}~
        \orcidlink{0000-0002-5589-9952}
        }
\else
\fi


\ifcase\authorStyle
    \author{R.~Thalmeier~\orcidlink{0009-0003-4480-0990}}
    \affiliation{\iMBI}
\or
    \author[\MBI]{
        \fnm{R.}~
        \sur{Thalmeier}~
        \orcidlink{0009-0003-4480-0990}
        }
\else
\fi

\ifcase\authorStyle
    \author{C.~Tomei}
    \affiliation{\iINFNRoma}
\or
    \author[\INFNRoma]{
        \fnm{C.}~
        \sur{Tomei} 
        }
\else
\fi

\ifcase\authorStyle
    \author{L.~Valla~\orcidlink{0009-0003-7140-9196}}
    \affiliation{\iMBI}
\or
    \author[\MBI]{
        \fnm{L.}~
        \sur{Valla}~
        \orcidlink{0009-0003-7140-9196}
        }
\else
\fi

\ifcase\authorStyle
    \author{M.~Vignati~\orcidlink{0000-0002-8945-1128}}
    \affiliation{\iSapienza}
    \affiliation{\iINFNRoma}
\or
    \author[\Sapienza, \INFNRoma]{
        \fnm{M.}~
        \sur{Vignati}~
        \orcidlink{0000-0002-8945-1128}
        }
\else
\fi

\ifcase\authorStyle
    \author{M.~Vivier~\orcidlink{0000-0003-2199-0958}}
    \affiliation{\iCEA}
\or
    \author[\CEA]{
        \fnm{M.}~
        \sur{Vivier}~
        \orcidlink{0000-0003-2199-0958}
        }
\else
\fi


\ifcase\authorStyle
    \author{A.~Wallach~\orcidlink{ 0009-0009-1703-9634}}
    \affiliation{\iTUM}
\or
    \author[\TUM]{
        \fnm{A.}~
        \sur{Wallach}~
        \orcidlink{0009-0009-1703-9634}
        }
\else
\fi

\ifcase\authorStyle
    \author{P.~Wasser~\orcidlink{0009-0004-7650-7307}}
    \affiliation{\iTUM}
\or
    \author[\TUM]{
        \fnm{P.}~
        \sur{Wasser}~
        \orcidlink{0009-0004-7650-7307}
        }
\else
\fi

\ifcase\authorStyle
    \author{A.~Wex~\orcidlink{0009-0003-5371-2466}}
    \affiliation{\iTUM}
\or
\author[\TUM]{
    \fnm{A.}~
    \sur{Wex}~
    \orcidlink{0009-0003-5371-2466}
}
\else
\fi

\ifcase\authorStyle
    \author{L.~Wienke~\orcidlink{0009-0006-5548-2109}}
    \affiliation{\iTUM}
\or
    \author[\TUM]{
        \fnm{L.}~
        \sur{Wienke}~
        \orcidlink{0009-0006-5548-2109}
        }
\else
\fi

\ifcase\authorStyle
\or


    \affil[\TUW]{\iTUW}

    \affil[\MPP]{\iMPP}
    
    
    \affil[\MBI]{\iMBI}

    \affil[\INFNTorVergata]{\iINFNTorVergata}

    \affil[\CEA]{\iCEA}

    \affil[\INFNRoma]{\iINFNRoma}

    \affil[\Sapienza]{\iSapienza}

    \affil[\TUM]{\iTUM}

    \affil[\TorVergata]{\iTorVergata}

    \affil[\Ferrara]{\iFerrara}

    \affil[\INFNFerrara]{\iINFNFerrara}

    

    %


    \newcommand{\symAlsoAtCoimbra}{\dag}
    \affil[\symAlsoAtCoimbra]{\iAlsoAtCoimbra}
    
    
    \newcommand{\symNowAtMPIK}{\ddag}
    \affil[\symNowAtMPIK]{\iNowAtMPIK}

    \newcommand{\symNowAtLNGS}{\#}
    \affil[\symNowAtLNGS]{\iNowAtLNGS}    


    \newcommand{\symNowAtKiutra}{\textdollar}
    \affil[\symNowAtKiutra]{\iNowAtKiutra}

    
     
\else
\fi

\abstract{
The NUCLEUS experiment aims to detect coherent elastic neutrino–nucleus scattering of reactor antineutrinos using low-threshold, gram-scale cryogenic calorimeters. Similar to other low-threshold experiments, NUCLEUS observes a sharp rise in the event rate below a few hundred eV, referred to as the low energy excess (LEE), whose origin remains yet unidentified. Building on results from the NUCLEUS testing and commissioning at the Technical University of Munich and from previous characterization campaigns, we present a comprehensive study of the background rate measured with a sapphire detector equipped with two transition-edge sensors under various experimental conditions. We find no evidence for a dependence of the LEE rate on the particle background level, whereas the results indicate that slower cooling-down procedures lead to lower initial LEE rates. The behavior of the LEE rate during the same cooldown is comparable across the measurements and is best described by a power law with a common exponent across datasets of $(-0.59\pm0.06)$, when time is expressed from the moment the detector reaches the \unit[4]{K} temperature. These findings provide valuable guidance for future LEE mitigation strategies in the NUCLEUS experiment.
}
    
\maketitle

\section{\label{sec:intro}Introduction}
Coherent elastic neutrino--nucleus scattering (\cevns[)] is a process in which a neutrino scatters off an entire nucleus via the weak neutral current~\cite{Freedman:1974,barranco,lindner,cevnsreview}. This process was first observed by the COHERENT experiment~\cite{coherent1} in 2017 with neutrinos from a spallation neutron source, followed by a series of measurements exploring the dependence of the process on the target nucleus~\cite{coherent2,coherent3}. More recently, an observation of \cevns with reactor antineutrinos was reported by the CONUS+ experiment~\cite{conus}, while the MINER~\cite{miner}, $\nu$GEN~\cite{nugen}, TEXONO~\cite{texono}, CONNIE~\cite{connie} and RED-100~\cite{red100} Collaborations presented upper limits on the \cevns cross-section. In addition, the Ricochet~\cite{Ricochet:NTL} experiment is currently taking data for a reactor \cevns measurement.

The NUCLEUS experiment aims to measure this process using reactor antineutrinos from the Chooz nuclear power plant~\cite{NucleusPRD:2017,NucleusEPJC:2017,NucleusChooz:2019}. With typical energies of a few MeV, they produce nuclear recoils below a few hundred eV in the \cawo and \alo\ targets. To achieve sensitivity to such small energies, NUCLEUS employs gram-scale cryogenic calorimeters equipped with tungsten Transition-Edge Sensors (TES), reaching energy thresholds of a few tens of eV, unprecedented among current \cevns experiments. This low threshold enables \cevns observation in the fully coherent regime and provides access to a significantly larger fraction of the reactor antineutrino spectrum compared to experiments with higher energy thresholds.

While TES-based detectors provide the required sensitivity, the \cevns region of interest is strongly affected by a low energy background of yet unknown origin~\cite{EXCESS_review:2022,EXCESS_review:2025}. This spectral feature, commonly referred to as the low energy excess (LEE), becomes dominant below several hundred eV and increases steeply toward lower energies. Similar behavior has been observed in many low threshold cryogenic detectors with a striking shared feature of the time-dependent decay of the event rate~\cite{EXCESS_review:2025, CRESST_LEE:2023}. Since the LEE currently constitutes the main factor limiting sensitivity not only for cryogenic \cevns experiments but also for light dark matter searches~\cite{EXCESS_review:2022,EXCESS_review:2025}, its study and mitigation have become central to the R\&D efforts of the cryogenic detector community, directly influencing experimental strategies~\cite{CRESST_2TES:2024,TESSERACTTwoChannel:2025,Ricochet:NTL,Supercdms,TESSERACT_stress:2024}. Reading out crystal targets with two sensors simultaneously---an approach pioneered by the CRESST~\cite{CRESST_2TES:2024} and TESSERACT~\cite{TESSERACTTwoChannel:2025} Collaborations---has proven effective in rejecting TES-related LEE events through the identification of non-coincident signals. However, a large fraction of the LEE is observed in coincidence between the two sensors, as expected for events originating from energy depositions in the crystal bulk, such as those from \cevns interactions or dark matter, which produce signals shared between the two channels. Various solid-state detector-related effects are currently being discussed as possible origins of this LEE component~\cite{EXCESS_review:2025}.

In this work, we present a systematic study carried out by the NUCLEUS Collaboration during the testing and commissioning phases at the Technical University of Munich (TUM). The study uses the same double TES detector, closely matching the one intended for operation at the reactor site, and investigates its performance under different conditions to gain new insights into the origin and behavior of the LEE. After introducing the experimental setup (Section~\ref{sec:setup}), datasets (Section~\ref{sec:runs}), and analysis strategy (Section~\ref{sec:ana}), we investigate potential factors affecting the measured LEE rates and their evolution in time, including particle backgrounds (Section~\ref{sec:bkg}) and cooldown parameters (Section~\ref{sec:rate_vs_time}).

\section{\label{sec:setup}Experimental Setup}
In this paper, we will focus on the results obtained with a sapphire (\alo[)] double TES detector operated at TUM in both an above-ground facility and a shallow underground laboratory. Here we discuss the setup features common to all the considered measurements, since both the shield against external radioactivity and the stress exerted on the detector by the mounting~\cite{TESSERACT_stress:2024} affect the measured counting rate.

\begin{figure}
    \centering
        \includegraphics[width=1.04\linewidth]{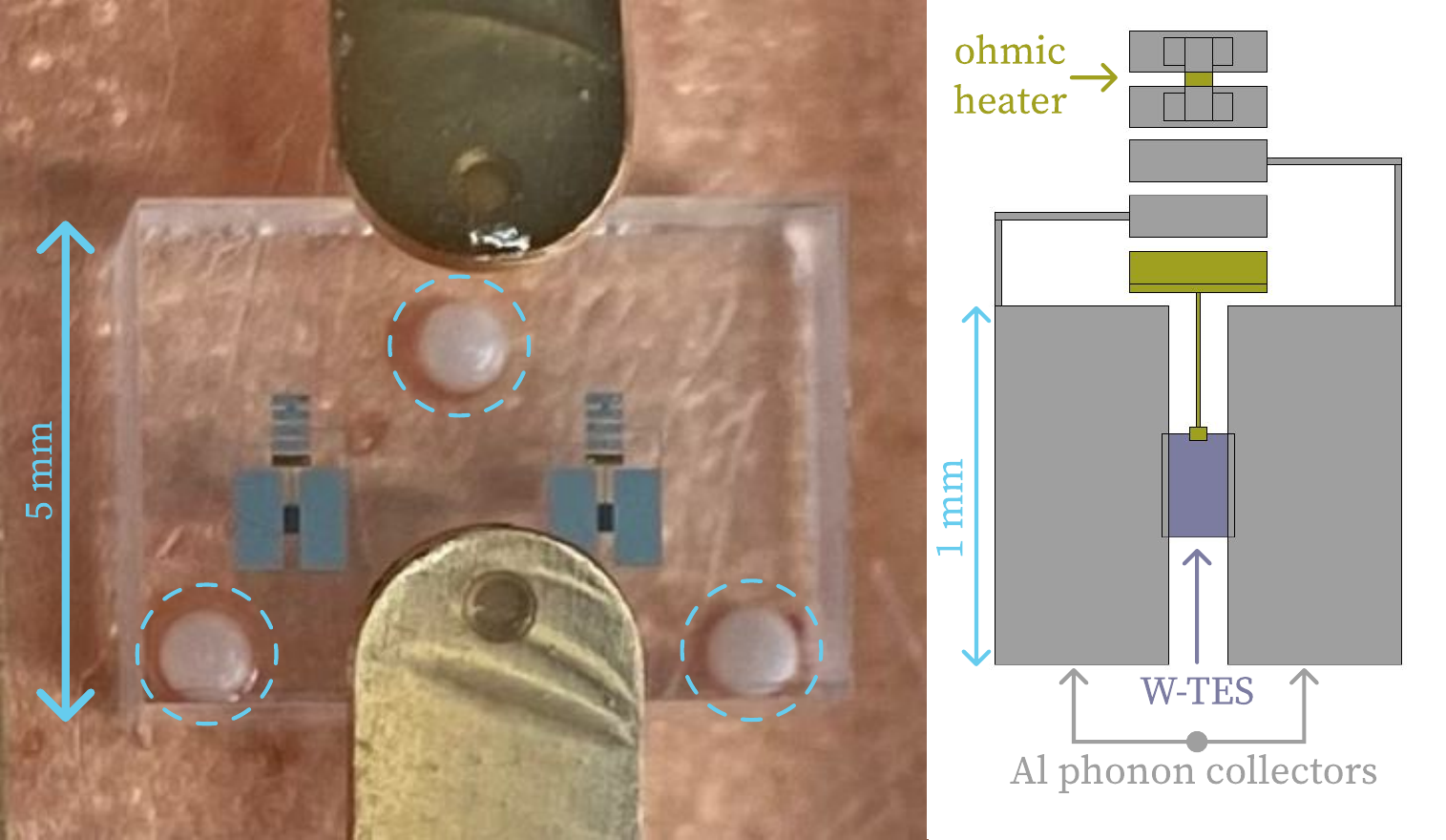}
        \caption{\textbf{Left:} Photograph of the double TES module. The \unit[0.75]{g} \alo\ crystal absorber, instrumented with two TESs, is mounted inside a copper housing. The detector is supported from below by three sapphire spheres of \unit[1]{mm} diameter (indicated by the blue dashed circles) and secured from above by two bronze clamps, each resting on an additional sapphire sphere. \textbf{Right:} Schematic of the TES used in the NUCLEUS double TES detector. The purple color represents the tungsten layer, the gray the aluminum and the yellow the gold. The tungsten and aluminum structures have a small overlap and are evaporated onto an auxiliary $\mathrm{SiO_2}$ layer between the crystal and the films.}
    \label{fig:dbl2}
\end{figure}

\paragraph{Detector Description} 
The detector consists of a \alo absorber with dimensions of $\unit[5\times5\times7.5]{mm^3}$ (\unit[0.75]{g}), shown in Figure~\ref{fig:dbl2}, instrumented with two TESs. Each sensor consists of a thin tungsten film slightly overlapping with aluminum phonon collectors and is thermally coupled to the thermal bath via a gold link. Both sensors, characterized by transition temperatures of \unit[15.1]{mK} and \unit[15.6]{mK} respectively, are biased to their optimal operating points within the superconducting transitions using one of the gold ohmic heaters evaporated onto the crystal surface and individual bias currents. The signals from the TESs are read out using commercially available Superconducting Quantum Interference Devices (SQUIDs) and are recorded as continuous data streams for each TES by the Versatile Data Acquisition (VDAQ).

The crystal is supported from below by three sapphire spheres of \unit[1]{mm} diameter and held in place by two bronze clamps. Each of the clamps has a sapphire sphere underneath to ensure a point-like contact to the crystal. The entire assembly is mounted inside a copper housing to shield against external thermal irradiation.

\paragraph{General setup}
The detector was operated in two experimental setups at the Physics Department of TUM. Both setups employ a BlueFors LD400 dry dilution refrigerator, achieving a base temperature of approximately \unit[10]{mK}. Pulse-tube and environmental induced vibrations are effectively attenuated by a dedicated vibration decoupling system, based on a cryogenic spring pendulum~\cite{WexSpring:2025}.  

The first setup, labeled \textit{Surf}, is located on the top floor of a two-story building and does not include any dedicated shield against background radiation. The second setup is located in the shallow underground laboratory (\textit{UGL})~\cite{Langenkamper2018ugl}, with an overburden of $\sim$\unit[10]{m.w.e.} \cite{LBR_paper}. The UGL cryostat is surrounded by an external passive shield consisting of \unit[5]{cm} of low-radioactivity lead (Pb) and \unit[20]{cm} of 5\% borated polyethylene (PE), mounted on a movable steel frame. This design allows for opening the shield, enabling easy access to the cryostat.  

\begin{figure}
    \centering
        \includegraphics[width=1\linewidth]{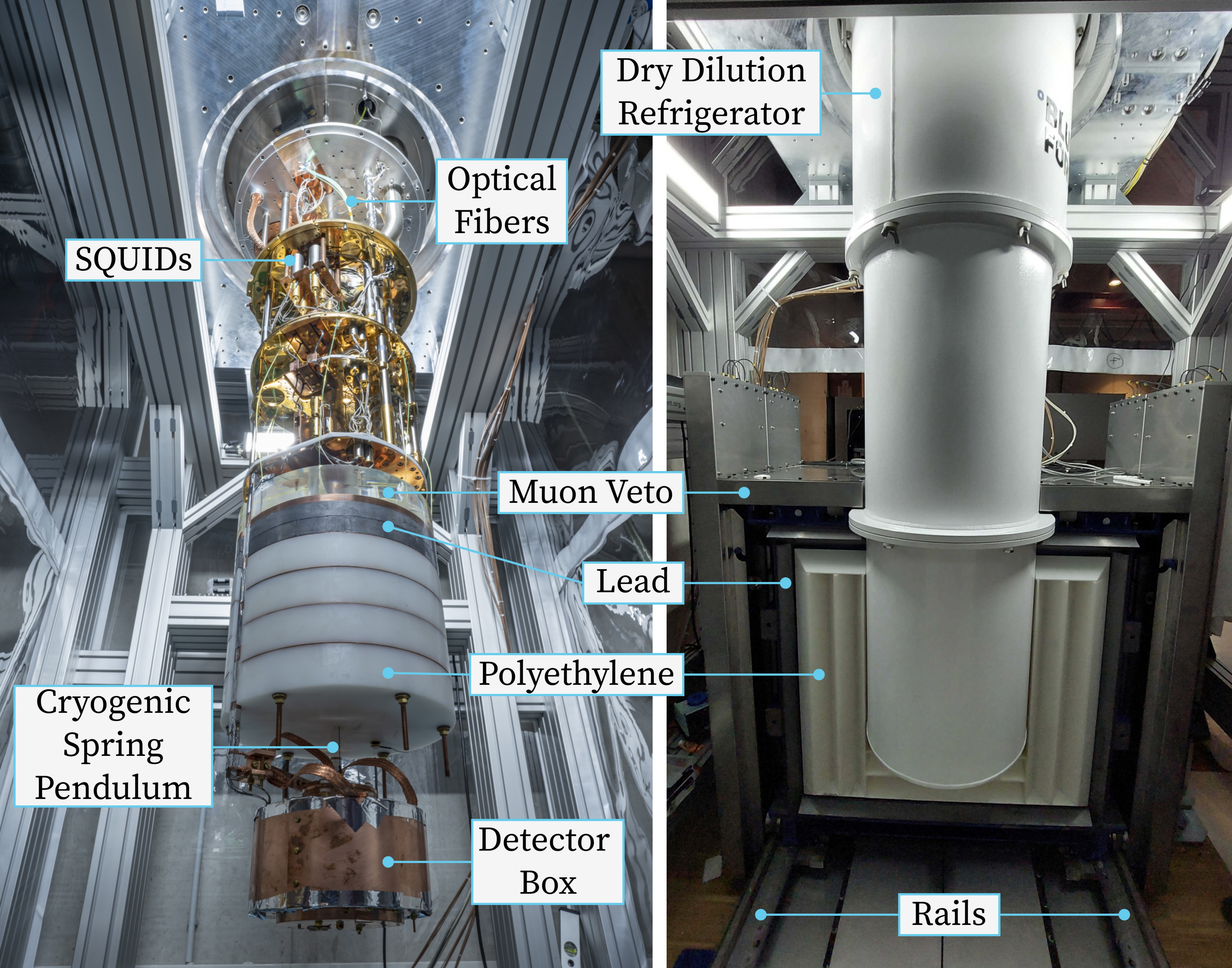}
        \caption{Photographs of the NUCLEUS commissioning setup at TUM deployed in the UGL experimental site, with the cryostat open (left) and closed (right). 
        Inside the cryostat, the detector box is mounted on a cryogenic spring pendulum serving as the vibration decoupling system \cite{WexSpring:2025}. The detectors are read out by SQUIDs, and optical fibers provide the LED optical calibration \cite{DelCastello_LANTERN:2024,lantern2}.
        The shield (PE, Pb, and active MV layers) is installed inside and outside the cryostat, with the external layers assembled into two half-cubes mounted on rails (in the right panel, one of the half-cubes is visible). Table~\ref{tab:measurements} indicates the shield configuration used in each of the considered measurements.
        }
    \label{fig:UGL_cryostat}
\end{figure}

In later measurements, the UGL setup, shown in Figure~\ref{fig:UGL_cryostat}, was further equipped with an active muon veto (MV) made of plastic scintillator panels read out by SiPMs~\cite{nucleus:MV}, surrounding the external shield. Additionally, a cylindrical inner shield was installed above the cryogenic detectors inside the cryostat to close the gap left by the external layers. Operated at sub-kelvin temperatures, it reproduced the same structure from the inside out: PE, Pb, and cryogenic MV~\cite{nucleus:cryoMV}. This configuration closely resembles the experimental setup planned for operation at the reactor site~\cite{nucleus_simulations} and is described in detail in~\cite{LBR_paper}.

\section{\label{sec:runs}Measurement Overview}
 
\begin{table*}
\centering
\caption{List of datasets used in this study (see Section~\ref{sec:runs}) presented in ascending chronological order; all data was acquired using the same \alo double TES detector. The \textit{Shield} column indicates the shield against environmental radiation used during data taking: the external shield (Ext.) is absent in the surface measurements and can be present in either open or closed configuration in the UGL measurements (see Figure~\ref{fig:UGL_cryostat} for shield details). The internal cryostat shield (Int.) was present only in the commissioning run. The last column provides the noise root mean square (RMS) $\sigma_{0}$ of the two TESs after applying the matched filter; the uncertainties for $\sigma_{0}$ reflect the spread between the values determined by the two independent analyses.}
\label{tab:measurements}
\resizebox{\textwidth}{!}{
\begin{tabular}{c|c|c|c|c|c|c } 

Label     & Cooldown & Location  & Shield & Duration (h)& Calibration & \makecell{Noise $\sigma_0$ (eV)\\TES 1\\TES 2}\\
\hline
\hline
Surf1        &  Surf1          &  Surface  &  No              & 12.0    & $^{55}$Fe & \makecell{$6.7\pm0.2$\\$6.1\pm0.2$}\\
&&&&&&\\
Surf2        &  Surf2          &  Surface  &  No              & 7.0     & \makecell{No\\(same as Surf1)} & \makecell{$6.6\pm 0.2$\\$7.1\pm0.7$}\\
&&&&&&\\
UGL1         &  UGL1           &  UGL      &  Ext. (closed)   & 109.4 & $^{55}$Fe & \makecell{$11.0\pm1.0$\\$8.4\pm0.4$}\\
\multicolumn{7}{c}{\cellcolor{gray!25}{Detector Remounting+Cleaning}} \\
&&&&&&\\
Surf3        &  Surf3          &  Surface  &  No              & 109.1 & $^{55}$Fe & \makecell{$6.5\pm0.8$\\$8.3\pm0.1$}\\
&&&&&&\\
UGL2-Closed  & \multirow{2}{*}[-5pt]{UGL2}
             & \multirow{2}{*}[-5pt]{UGL}
             & Ext. (closed)   & 64.1     & $^{55}$Fe & \makecell{$8.2\pm0.2$\\$6.1\pm1.0$}\\
UGL2-Open    &                 & 
             & Ext. (opened)   & 2.7      & $^{55}$Fe & \makecell{$11.9\pm1.0$\\$8.4\pm0.3$}\\
&&&&&&\\
Comm         &  Commissioning  &  UGL      &  \makecell{Int.+\\Ext. (closed)+\\Muon Veto} & 572.0 & LED & \makecell{$5.8\pm0.3$\\$5.5\pm0.2$}\\
\hline
\end{tabular}
}
\end{table*}

 In preparation for the NUCLEUS commissioning measurement~\cite{LBR_paper}, the same double TES detector was operated in multiple cryogenic R\&D runs at both the Surf and UGL setups during 2023 and 2024. In this work, we include all datasets from this period that exhibit satisfactory detector performance. A chronological overview of the datasets used in this study is given in Table~\ref{tab:measurements}. For convenience, the datasets are named according to the setup in which they were recorded, except for the full commissioning run, labeled \textit{Comm}.

In the Surf1, Surf2, and UGL1 datasets, the detector faced a commercially manufactured printed circuit board (PCB) used for sensor readout, which was found to contribute significantly to the particle background. For later measurements (Surf3, UGL2, and Comm), the PCB was replaced by custom copper–kapton–copper traces glued to a copper plate. In addition, the detector module was dismounted to clean and etch all components in the vicinity of the crystal, including the housing and clamps, to further reduce radioactive contamination.

For most measurements, a $^{55}$Fe source emitting \unit[5.5]{keV} and \unit[6.5]{keV} Mn K$_\alpha$ and K$_\beta$ lines was used for energy calibration. When present, the source was positioned above the detector, irradiating the TES side of the crystal. In the Surf2 dataset, however, the $^{55}$Fe source was removed, and the calibration factor obtained for Surf1 was used. For the Comm measurement, a more sophisticated approach was applied: the crystal was illuminated with LED pulses via a light fiber as described in~\cite{DelCastello_LANTERN:2024,lantern2}. The pulse intensity was controlled by the number of photons (each with an energy of $E_\gamma = \unit[4.86]{eV}$) per pulse. This method allowed for determining the detector response over a broad energy range using measured LED pulse amplitudes and photon statistics. The calibration strategy for the Comm dataset is detailed in~\cite{LBR_paper}.

The different background levels, resulting from the varying overburden of the two setups, surrounding materials, and shield configurations, enable LEE studies under diverse external conditions (Section~\ref{sec:bkg}). Furthermore, operating the same detector in different runs allows us to study the dependence of the LEE rate on various cooldown parameters (Section~\ref{sec:rate_vs_time}).

\section{\label{sec:ana}Analysis Description}

\begin{figure*}[t]
        \centering
        \includegraphics[width=1\linewidth]{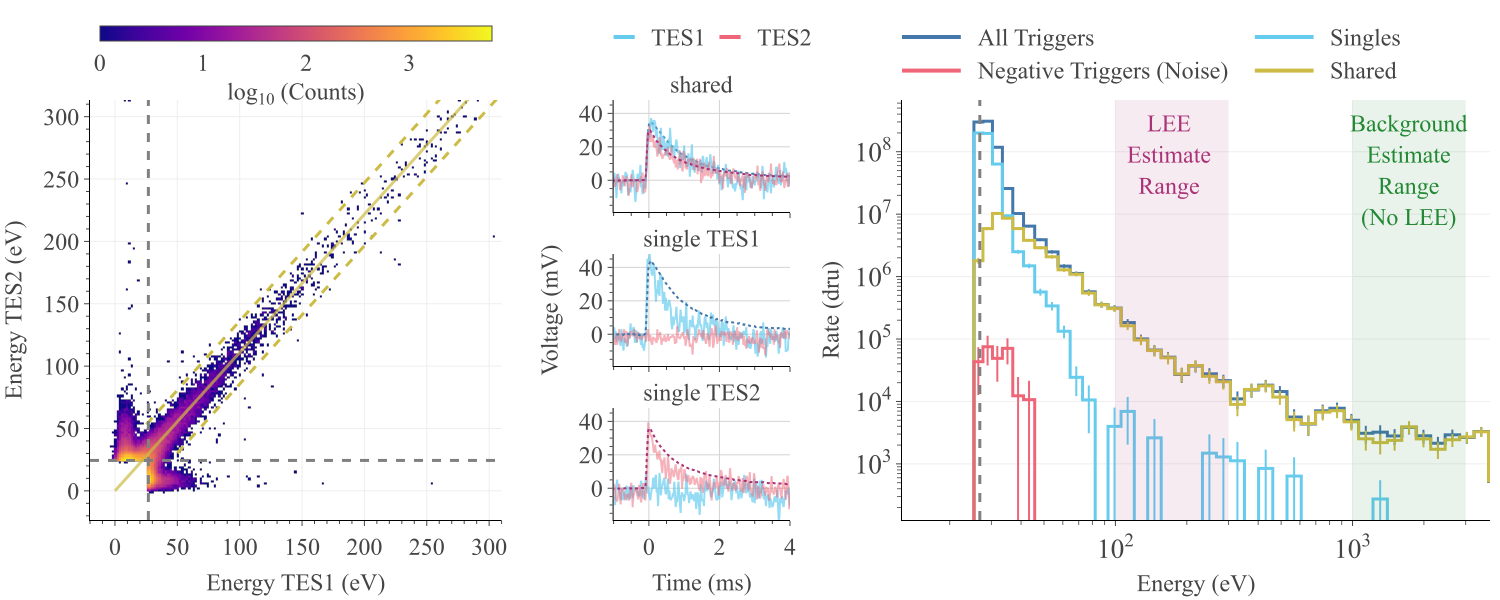}
        \caption{
        \textbf{Left:} Energy sharing between the two TES channels at low energies. Events between the yellow dashed lines, defined by Eq.~\ref{eq:shared_cut}, are classified as \textit{shared} events. Signals appearing in only one TES within the coincidence window are identified as \textit{singles}; the grey dashed lines indicate the triggering thresholds and define the \textit{singles}-event selection. The solid yellow line shows the fit $E_\text{TES2} = r\cdot E_\text{TES1}$ used to extract the parameter $r$, which accounts for calibration mismatches.
        \textbf{Center:} Example coincident traces at energies around \unit[100]{eV} from both TESs for the event populations identified in the left panel: \textit{shared} events, where both sensors record a pulse, and \textit{singles}, where only one sensor does. Typical pulse shapes from higher-energy particle events are shown as dashed lines. 
        \textbf{Right:} Energy spectra, in dru ($\unit[1]{dru} = \unit[1]{counts/(keV\,kg\,d)}$), recorded during the NUCLEUS commissioning run \cite{LBR_paper}, showing all triggered events (blue), shared events (yellow), singles (cyan), and negative triggers used to estimate the noise-induced trigger rate (red). The magenta shaded area marks the $[100,300]\,\mathrm{eV}$ interval used for the LEE estimate, while the green shaded area indicates the region used to evaluate the background $\mathrm{Rate}([1,3]\,\mathrm{keV})$. The vertical dashed gray line denotes the triggering threshold at $4\sigma_0$.}
        \label{fig:energy_spectra}
\end{figure*}

The analysis of the data recorded with the two TESs of the detector involves several steps designed to extract meaningful features from the raw data stream. The two TES channels are first analyzed independently, after which their results are combined to classify the events. 

The first step is the offline triggering of events. This is performed by applying a threshold trigger algorithm on the data stream filtered with a matched filtering technique to maximize the signal-to-noise ratio (SNR)~\cite{optimum_trigger}. The filter is built based on the characteristic signal pulse shape and the noise observed in a given measurement. It is also used to evaluate the detector’s baseline resolution, $\sigma_0$, characterized by the RMS of the filtered noise. The values of $\sigma_0$ for all measurements presented are listed in Table~\ref{tab:measurements}. These are reported in eV using the calibration constant extracted following the procedure described in the next paragraph. Once the signal in the filtered data stream exceeds $4\sigma_0$ for any of the TESs, the data traces from both sensors are selected around the trigger position for subsequent coincident analysis.

The second step in the analysis is to evaluate several basic pulse shape parameters, as described in \cite{NUCLEUS_diana_GDC, cait}, carried out separately for each TES. These parameters are used to clean the data from electronic glitches, pile-ups, and saturated signals. In this way, a clean selection of pulses is identified. The amplitude of the pulses can then be reconstructed by taking the maximum of the filtered signals and used as an estimation of the energy deposited in the detector. The measure of the amplitude is calibrated as an energy response using either the X-ray peaks produced by $^{55}$Fe or an LED optical calibration~\cite{DelCastello_LANTERN:2024,lantern2}, depending on the dataset under consideration, as listed in Table~\ref{tab:measurements}. The discrepancy between these two calibration methods has been measured during the NUCLEUS commissioning~\cite{LBR_paper} to be lower than 25\%, which is subdominant with respect to the observed variations in the counting rate presented in this work. The amplitude reconstruction exploits the fact that the energy deposited in the crystal is determined from the height of the measured pulse. As long as the deposited energy remains within the linear response range of the detector, the pulse shape stays constant and is well described by the model developed in~\cite{PulseShapeModel1995}, thus matched filtering techniques can be applied. Figure~\ref{fig:energy_spectra} (left) shows the reconstructed energies of all clean events in the two sensors from the Comm dataset, forming distinct event populations explained in the following paragraphs.

Due to the presence of two sensors on the same crystal, a single energy deposition in the absorber is measured by both TESs.  This double readout enables the classification of events into two categories: \textit{shared} events, occurring in two sensors simultaneously within a time window of $\pm$\unit[0.5]{ms} and reconstructed with compatible energies in both TESs, and \textit{singles}, events triggered by only one of the two sensors within this time window. During the Comm measurement, LED-induced calibration pulses were generated and seen to populate the diagonal band of Figure~\ref{fig:energy_spectra} (left), confirming approximately equal energy reconstruction between the two sensors for energy depositions in the bulk of the crystal substrate. In addition to the two main classes of events considered in this work, events with a hybrid classification can also be found in the data where two signals are present but do not show the expected amplitude relation. This family of events is discarded from this work since they are assumed to be due to position effects (e.g. TES direct hits) or working point fluctuations.

Examples of the \textit{shared} and \textit{singles} event classes are shown in the middle panel of Figure~\ref{fig:energy_spectra} and compared to a template pulse shape constructed from particle events in the energy range from 0.5 to \unit[3]{keV}. These reference events are known to occur inside the crystal, are measured by both sensors, and therefore belong to the shared band. While low energy events in the shared band closely follow this template pulse shape, \textit{singles} appear slightly faster in both channels. Despite this observed difference in pulse shape, the classification of \textit{shared} events is based on a comparison of the energies measured by the two TESs, $E_{\mathrm{TES1}}$ and $E_{\mathrm{TES2}}$, which provides a more robust discrimination criterion, particularly at low energies where pulse-shape differences are no longer reliably resolved.

In an ideal case, the events would align along the diagonal $E_\mathrm{TES2} = E_\mathrm{TES1}$. However, since the two sensors are calibrated independently and may exhibit different non-linearity effects, small mismatches of up to 10\% are observed. To account for these effects, we fit a linear function $E_\mathrm{TES2} = r \cdot E_\mathrm{TES1}$ to the event distribution, as shown by the solid yellow line in Figure~\ref{fig:energy_spectra} (left). Under the assumption that the corrected energy difference is mainly due to baseline noise, the \textit{shared}-event selection cut is applied as:
\begin{equation}\label{eq:shared_cut}
|E_\mathrm{TES2} - r \cdot E_\mathrm{TES1}| < 3 \cdot \sqrt{\sigma^2_{0,\mathrm{TES1}} + r^2\cdot\sigma^2_{0,\mathrm{TES2}}},
\end{equation}
where $\sigma_0$ denotes the noise RMS of each TES. As visible from the equation above, the \textit{shared} band selection is performed in a $\pm 3$ sigma band around the energy difference of the two sensors. This interval was chosen in order to retain most events at higher energies while avoiding major contamination from the \textit{singles} in the lowest energy bins. The resulting selection region is illustrated by the dashed yellow lines in Figure~\ref{fig:energy_spectra} (left).

The analysis efficiency, which accounts for the combined effects of triggering, reconstruction, and data cleaning, is evaluated using simulated pulses. 
The analysis procedure is applied to these pulses, which have a characteristic shape and are injected into the data at various amplitudes to cover the entire energy range under consideration. In the absence of a control population in most datasets, the efficiency for the \textit{shared}/\textit{singles} band selection is assumed to be 100\% within the quoted systematic uncertainty. The discrimination power of this selection, i.e., its ability to distinguish \textit{shared} events from \textit{singles}, inevitably degrades at the lowest energies. However, as discussed later in this study, we focus on the energy range above \unit[100]{eV}, where a reliable discrimination between the \textit{shared} and \textit{singles} populations is achievable in all datasets, justifying this choice. The fraction of events between \unit[500]{eV} and \unit[1]{keV} that lie outside the \textit{shared} band is used as a systematic uncertainty on the efficiency, as this energy range is not dominated by the LEE and such events are expected to originate from energy deposits in the crystal and thus should fall within the \textit{shared} band. This uncertainty is at the level of about 10\% for all datasets and is therefore comparable to the uncertainty on the energy calibration constant, which directly impacts the level of the considered spectra.

Since the signals from the TESs appear as sharp upward pulses in the data stream, the same triggering technique described above can be applied with a negative threshold to detect downward fluctuations. These negative fluctuations arise solely from the noise affecting the sensors. As the noise oscillates symmetrically around the signal baseline, such \textit{negative triggers} provide an estimate of the rate of false positive triggers in the data.
The resulting spectra, showing both the shared and singles populations, are displayed, in dru ($\unit[1]{dru} = \unit[1]{counts/(keV\,kg\,d)}$), in the right panel of Figure~\ref{fig:energy_spectra}. No correction is applied to correct the spectra for the negative triggers distribution, shown in red in Figure~\ref{fig:energy_spectra} (Right).

The described analysis is performed independently for each of the datasets listed in Table~\ref{tab:measurements}. The only measurement with small variation in the analysis protocol is the one performed for the commissioning run \cite{LBR_paper}, where LED-induced pulses were used in place of simulated signals. All datasets were analyzed with two different analysis software packages: Diana, a C++/Python~3 framework originally developed for the CUORE/CUPID experiments \cite{cuore,cupid}, adapted to NUCLEUS \cite{NUCLEUS_diana_GDC} and also employed in BULLKID~\cite{Delicato:2024adb}; Cait, a Python 3 package designed for processing raw data from cryogenic particle detectors \cite{cait} used in the CRESST~\cite{cresst}, COSINUS~\cite{cosinus}, and NUCLEUS experiments. Both analyses followed two similar but independent pipelines and yielded measured energy spectra consistent within 3$\sigma$, where $\sigma$ is defined through the significance of the difference, i.e.\ the difference between the two spectra divided by the quadratic sum of their uncertainties. Due to the good compatibility of the results produced by the two analyses, an average between the two has been used to present the results in this work, while the discrepancy between the analyses is taken as an uncertainty on the data points to better account for systematic effects, which dominate the errors of the presented results.

For each measurement, the energy spectrum is constructed using the worst-performing TES (i.e., the channel with the higher noise RMS), while the second channel is used for event identification. This choice was made in order to have a valid event classification down to the detector threshold. The spectral shape of the \textit{singles} component is steeper than that of the \textit{shared} LEE. Consequently, the \textit{singles} contribution becomes dominant at lower energies, with the exact energy at which this occurs depending on the relative rates of the two components and varying across different measurements (from $\sim\!\unit[100]{eV}$ in UGL1 to $\sim\!\unit[40]{eV}$ in the Comm run, while in the surface measurements this crossover is not observed within the accessible energy range due to their higher \textit{shared}-LEE rates).

For the purpose of this study, an estimate for the LEE rate is performed in the \unit[100-300]{eV} energy range, chosen for several reasons: the upper bound ensures sufficient statistics while remaining close to the region of interest for \cevns interactions. The lower bound is set well above the reconstructed energy of the highest negative-trigger event across all datasets, thereby excluding spurious noise contributions from the results. In addition, the \unit[100]{eV} threshold provides reliable discrimination between \textit{singles} and \textit{shared} events throughout all the datasets, making this a safe choice for the study.

The spectrum between 1 and \unit[3]{keV} is assumed to be flat and dominated by physical particle events and is therefore used as a particle background benchmark, following the approach of \cite{LBR_paper}. This background level serves as the reference above which the LEE develops.

Considering these two spectral regions, the LEE event rate is defined as:
\begin{equation}\label{eq:rlee}
    R_\text{LEE} = \mathrm{Rate}([100,300]\,\mathrm{eV})-\mathrm{Rate}([1,3]\,\mathrm{keV}),
\end{equation}
where $\text{Rate([\textit{A,B}])}$ is the measured rate, expressed in dru, between energies $A$ and $B$. Thus, this procedure does not rely on any assumption about the spectral shape of the LEE, but is intended to provide an estimate of the LEE rate independent of the particle background, which is assumed to be subdominant at low energies. As previously mentioned, particle interactions occurring in the crystal absorber, such as CE$\nu$NS, are characterized by simultaneous signals in both TESs with similar reconstructed energies. Accordingly, the studies presented in this work focus on events in the \textit{shared} band, which mimic this signature.

\section{\label{sec:bkg}Energy Spectra and correlation with physical backgrounds}

Due to differences in the experimental setups, varying background levels are observed across the datasets. This enables a systematic study of the LEE rate under different particle rate conditions, and the corresponding results are presented in this section.

    \subsection{\label{sec:particle_bkg}Effect of keV-Range Background on the LEE}
        \begin{figure}
            \centering
            \includegraphics[width=1.04\linewidth]{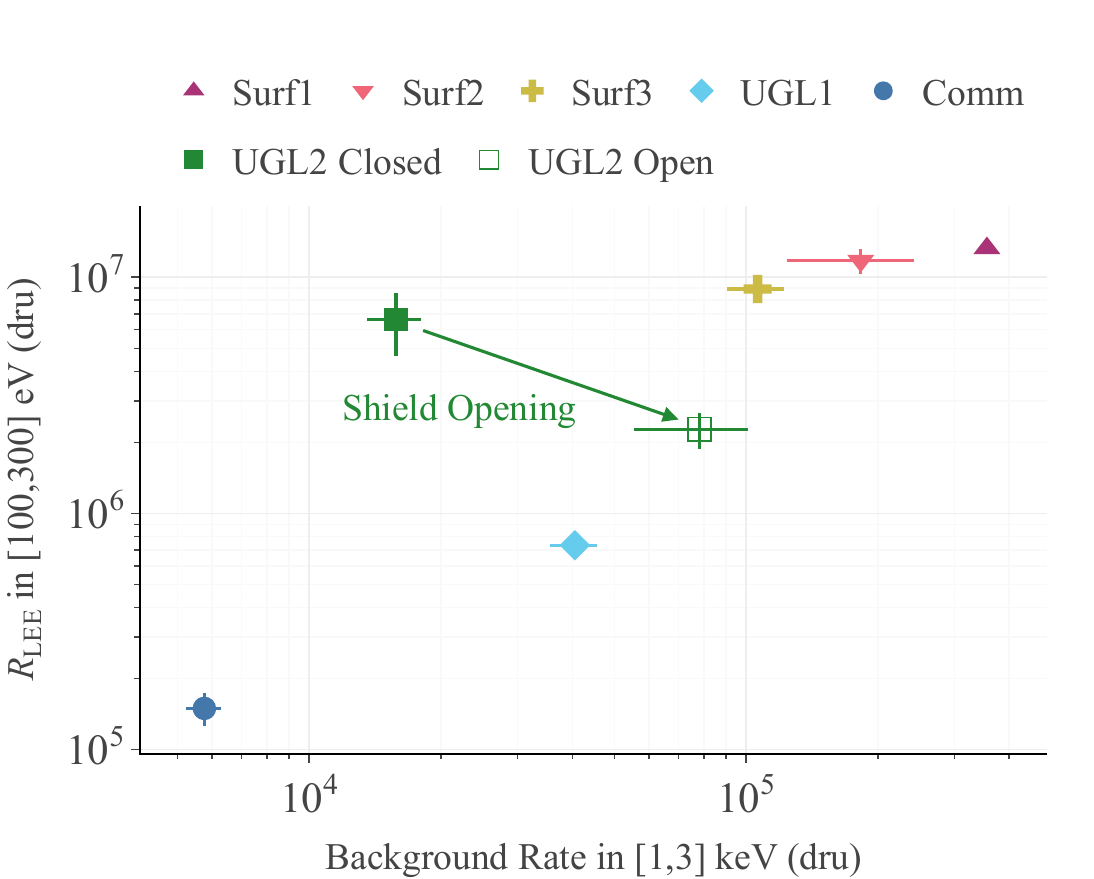}\\
            \includegraphics[width=1.04\linewidth]{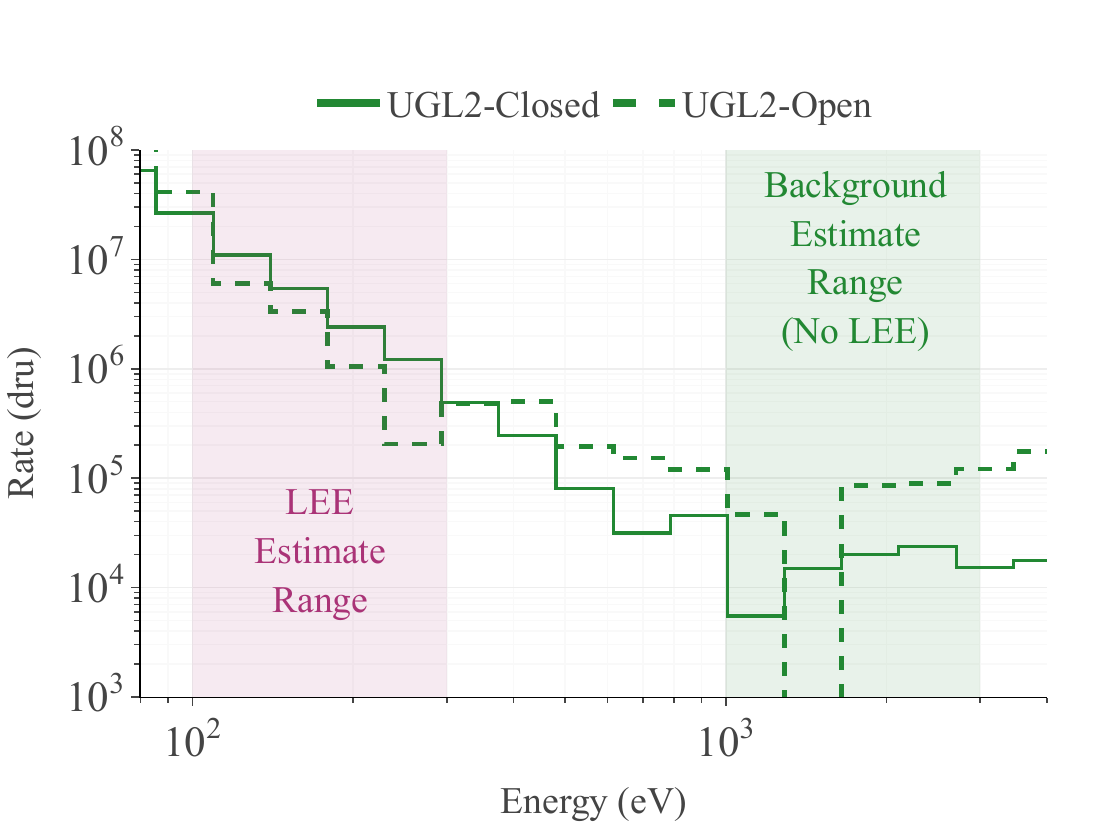} \\
            \caption{\textbf{Top:} Correlation of the LEE rate ($R_\text{LEE}$) with the background rate in the keV region for different datasets. A dedicated test during the UGL2 cooldown was performed by opening the shield surrounding the detector. Under the hypothesis of a background-induced LEE, the $R_\text{LEE}$ was expected to increase; the opposite was observed. \textbf{Bottom}: Energy spectra of the two UGL2 datasets, showing an increase in the particle background rate in the keV region (shaded in green) when the shield is opened, while the LEE region (shaded in pink) exhibits the opposite behavior.}
            \label{fig:background_corr}
        \end{figure}

        In this study, we will discuss the correlation between the background rate measured in the keV-range and $R_\text{LEE}$. For this purpose, we consider the event rate in the energy range from \unit[1]{keV} to \unit[3]{keV}, as an estimate of the overall particle background rate. As presented in~\cite{LBR_paper}, this region is reproduced by Geant4 simulations~\cite{nucleus_simulations} up to a factor 1.5 discrepancy, with the indication that this difference is due to a limitation in the absolute value of the Monte Carlo result and not related to the LEE.
        
        The top panel of Figure~\ref{fig:background_corr} presents the average LEE rate over the measurement period versus the background rate in the keV region for all datasets. From the plot it is visible how the Surf data points show a correlation between the keV-background and the LEE-background; a similar behavior can be noticed between the Comm and UGL1 runs. To confirm or disprove such a correlation, a dedicated test was performed with the UGL2 cooldown. Before the UGL2 run, the setup was thoroughly cleaned, resulting in a lower background rate than in UGL1, but this did not affect the LEE, which actually is higher in the UGL2 measurements. In addition, it is worth noting that the hadronic component of cosmic rays was strongly suppressed by operating in the underground UGL setup and by employing the NUCLEUS external shields; despite these conditions, the observed LEE rate in UGL2 was comparable to that measured at the surface. To further probe a possible background dependence a dedicated test during the UGL2 run was performed: after recording the UGL2-Closed dataset, the shields were opened recording the data labeled as UGL2-Open, which, as expected, shows a higher background in the keV region. This setup modification, however, did not lead to any increase in the LEE rate, which, instead, was observed to decrease, as shown in the bottom panel of Figure~\ref{fig:background_corr}. Therefore, this is the first strong evidence we present in this work that the LEE rate in the NUCLEUS detectors has no indication of being induced by background driven mechanisms.
        
        Two important considerations must be highlighted here to address the initial incorrect hypothesis of the  scaling of the LEE with the background rate. First, the LEE rate is known to decrease as a function of time since cooldown, an effect not accounted for in Figure~\ref{fig:background_corr} but analyzed in detail in Section~\ref{sec:rate_vs_time} and shown in Figure~\ref{fig:rate_vs_time}. In fact, if a correction for the time decay of the LEE is applied, anticipating the results of Section~\ref{sec:rate_vs_time}, the expected LEE rate for UGL2-Open is \unit[$(4.8\pm0.3)\cdot 10^6$]{dru}, which is compatible at the level of 2 standard deviations with the measured value of \unit[$(2.3\pm1.2)\cdot 10^6$]{dru}. This then explains the difference in the measured LEE rates between the UGL2 datasets, without requiring any hypothesis to regarding the evolution of the rate and the opening of the shield.
        
        Second, achieving lower background levels typically involves additional internal shield, which prolongs the cooldown duration and delays the start of the measurement. These factors introduce degeneracies between the particle background level and time-related effects. Taken together, the data does not suggest any dependence between the observed LEE rate and the ambient particle background in the NUCLEUS detector and setup. This conclusion is further supported by previous studies from the CRESST Collaboration conducted in both deep-underground~\cite{CRESST_LEE:2023} and surface environments~\cite{CRESST_2TES:2024}.

    \subsection{\label{sec:mv}Effect of the Muon Veto on the LEE}

    \begin{figure}
            \centering
            \includegraphics[width=1.04\linewidth]{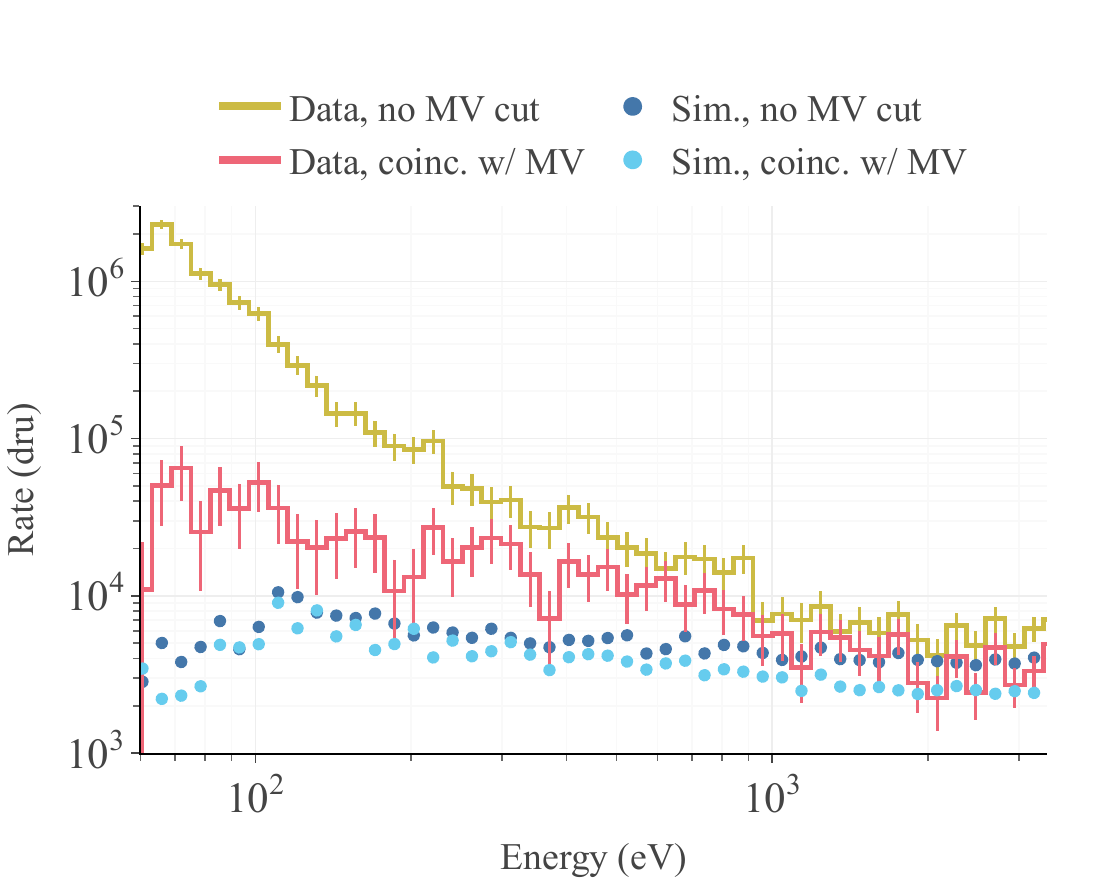}
            \includegraphics[width=1.04\linewidth]{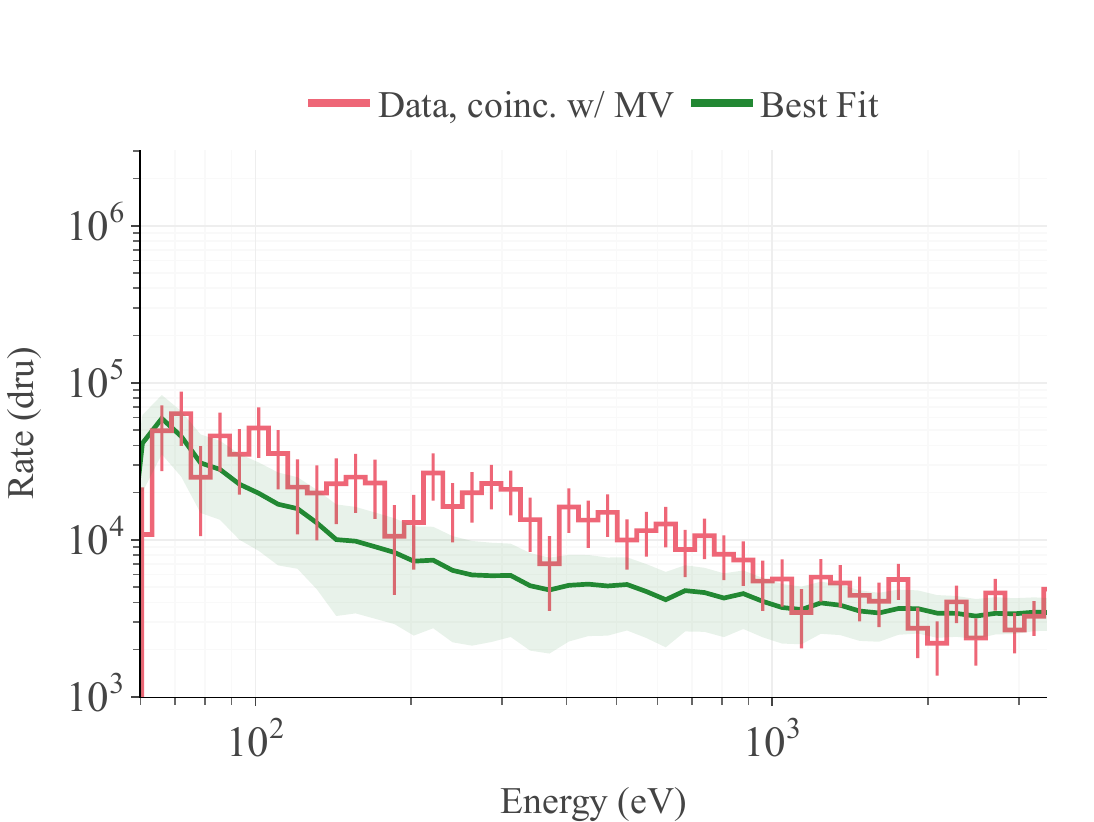}
            \caption{Effect of the MV coincidence cut measured during the NUCLEUS commissioning run \cite{LBR_paper}. \textbf{Top:} Measured energy spectra (solid histograms) before (yellow) and after (red) the application of the MV coincidence cut. The points show the Geant4 simulation results in the same range, before MV cut (blue) and after MV cut (cyan). \textbf{Bottom:} Best fit between toy model and data to extract the coincidence probability between the MV and the LEE spectrum. }
            \label{fig:mv_coinc_spec}
        \end{figure}
        
        In the commissioning run of the NUCLEUS experiment~\cite{LBR_paper} the MV was operated simultaneously with the cryogenic calorimeters. This gives the unique opportunity to study the effect that the high muon flux (\unit[0.0086]{muons/cm$^2$/s}) present in a shallow underground laboratory has on the low energy background. 

        The MV coincidence cut (abbreviated in the following to \textit{MV cut}) used for this analysis is described in \cite{LBR_paper}; in particular, for the detector under consideration it consists of keeping only the events that have a muon veto trigger in a $\pm \unit[60]{\mu\text{s}}$ time window near the trigger point of the TES signal. This time window gives an accidental coincidence probability of $\overline{p_\text{acc}}=(0.012\pm0.001)$, when evaluated on pulses injected via a transient excitation of the ohmic heater. The effect of the muon veto coincidence selection can be seen in the top panel of Figure~\ref{fig:mv_coinc_spec}, along with the corresponding Geant4 simulated energy spectra. As more thoroughly discussed in \cite{LBR_paper}, an overall normalization discrepancy of a factor $1.5\pm 0.1$ was observed between Monte Carlo and data above \unit[1.4]{keV}, while at low energies the presence of the LEE is not taken into account in the simulation. 

        With the data presented in Figure~\ref{fig:mv_coinc_spec}, along with the total simulated spectrum, it is possible to build a toy model in order to check whether the MV cut affects the LEE in a significant way. The toy model consists in assuming that the spectrum after the application of the MV cut can be separated in two components: the LEE  and the physical background that is dominated by real coincidences. Due to a lack of knowledge on the expected spectral shape of the LEE a similar approach to the one described in Section~\ref{sec:ana} is used to define this component. The LEE spectrum $S_\text{LEE}(E)$ is defined as the difference between the total measured spectrum $S^\text{tot}_\text{data}(E)$ and the simulated spectrum $S^\text{tot}_\text{simu}(E)$ rescaled by a parameter $\alpha$ that accounts for the spectral normalization discrepancy:
        \begin{equation}
            S_\text{LEE}(E)= S^\text{tot}_\text{data}(E)-\alpha S^\text{tot}_\text{simu}(E),
        \end{equation}
        where $\alpha$ is fixed to $\overline{\alpha}=1.5$ which is the discrepancy factor found in \cite{LBR_paper}. This value of  $\overline{\alpha}$ was found by comparing the data in the $\left[1.4,3\right]\unit{keV}$ energy range with simulations. In order to not inject twice the information of this data in the toy model, only events with energies below \unit[1.4]{keV} will be considered for this discussion.
        
        The toy model can then be built to test the assumption that the particle induced background, modeled as $\overline{\alpha} S^\text{tot}_\text{simu}(E)$, is vetoed with a probability equal to the one measured from real coincidences happening at higher energies $\overline{p_\text{MV}}=(0.62\pm0.05)$ and $S_\text{LEE}(E)$ is vetoed with probability $p_\text{LEE}$. The measured vetoed spectrum can then be modeled as:
        \begin{equation}
        \begin{aligned}
        S_{\text{LEE}}^{\text{MV}}(E \mid \epsilon_{\text{LEE}})
        &= p_{\text{LEE}}\, S_{\text{LEE}}(E)
           + \overline{p_{\text{MV}}}\,\overline{\alpha}\,
             S_{\text{simu}}^{\text{tot}}(E) \\
        &= p_{\text{LEE}}
           \left(S_{\text{data}}^{\text{tot}}(E)
           - \overline{\alpha}\, S_{\text{simu}}^{\text{tot}}(E)\right) \\
        &\quad
           + \overline{p_{\text{MV}}}\,\overline{\alpha}\,
             S_{\text{simu}}^{\text{tot}}(E),
        \end{aligned}
        \end{equation}
        where $p_\text{LEE}$ is the free parameter of the model. By fitting this model to the measured coincident energy spectrum of  the bottom panel of Figure~\ref{fig:mv_coinc_spec} the value of $p_\text{LEE}$ can be extracted and is estimated to be:
        \begin{equation}
            p_\text{LEE} = 0.025 \pm 0.004.
        \end{equation}
        This value of the muon coincidence probability of the LEE is at the same level of the accidental one $\overline{p_\text{acc}}=(0.012\pm0.001)$, suggesting that this spectral component is mostly affected by random coincidences, while muon-induced features are subdominant. The disagreement between these values can be explained by the simplicity of the used approach. In fact, from the bottom panel of Figure~\ref{fig:mv_coinc_spec}, it is visible that in the few hundred eV region, the agreement between the toy model and the data is not optimal. This could also be due to a worsening of the accuracy of the Monte Carlo in this energy range. On the other hand, satisfactory agreement with the measured data is found at the edges of the presented spectrum: at energies above \unit[1]{keV} the coincidences are dominated by muon crossings while at energies below \unit[100]{eV} they are mostly accidental. 

        Within the sensitivity of this order-of-magnitude study, this result shows no evidence that the LEE rate is affected by muon crossings beyond percent level, implying that more than 98\% of the measured LEE has a different origin.

\section{\label{sec:rate_vs_time}LEE rate evolution with time and impact of the cooldown parameters}

\begin{figure}
        \centering
        \includegraphics[width=1\linewidth]{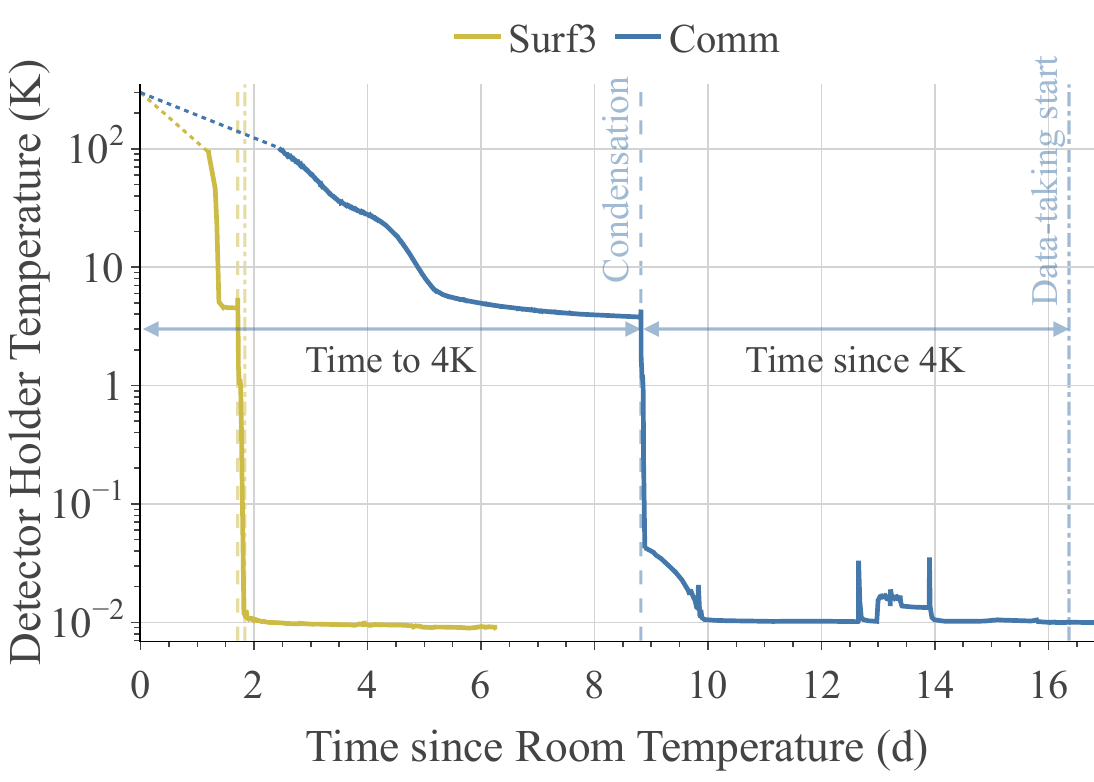}
        \caption{Temperature measured by a thermometer installed on the detector holder during the cooldown of the Surf3 and Comm runs. Time is shown relative to the start of cooldown from room temperature, defined by switching on the pulse tube. The thermometer begins recording once the temperature reaches \unit[100]{K}. The time of the relevant \unit[4]{K} condensation temperature is shown with vertical dashed lines; this moment is taken as zero on the x-axis of Figure~\ref{fig:rate_vs_time}. The dash-dotted lines mark the beginning of data taking. The evolution and duration of the different cooldown curves strongly depend on the cryostat configuration and on the mass and material of the mounted payload, therefore it can vary significantly across runs as shown in this plot.}
        \label{fig:cooldown_curve}
\end{figure}

\begin{figure*}[t]
        \centering
        \includegraphics[width=1\linewidth]{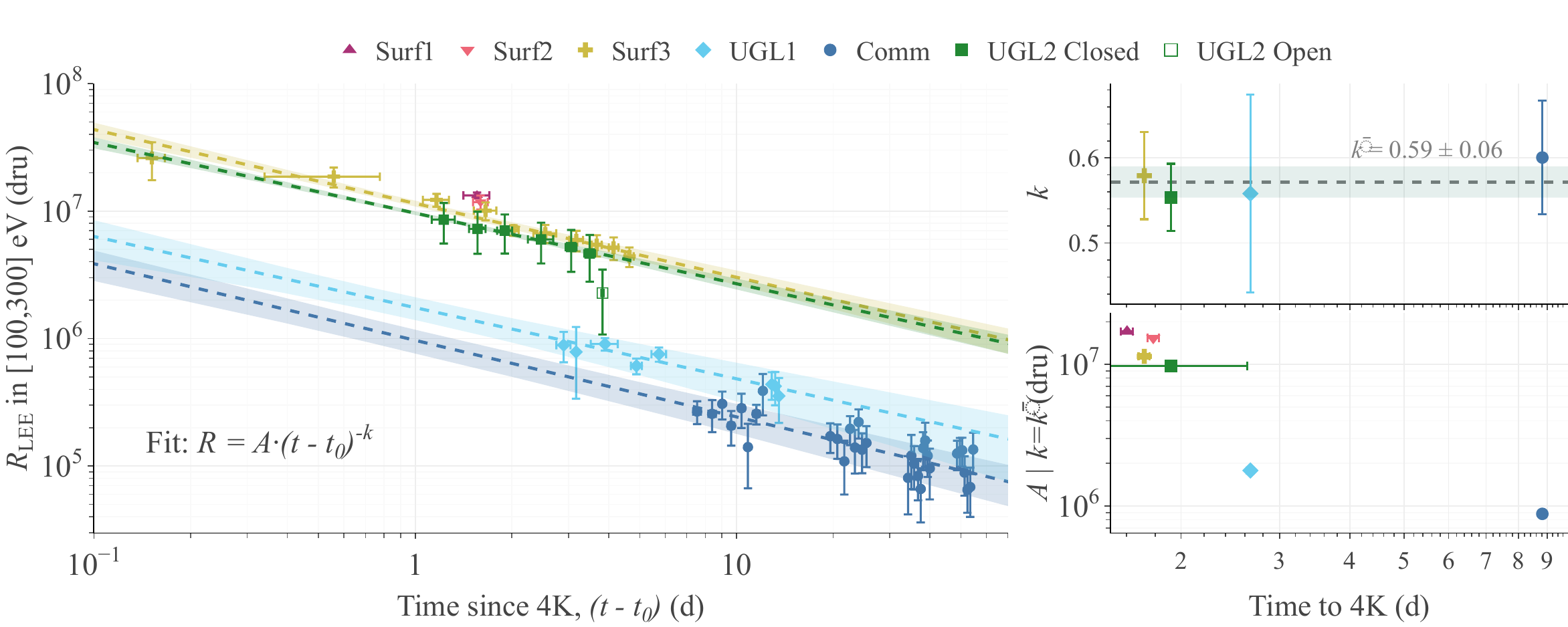}
        \caption{\textbf{Left:} LEE rate $R_\mathrm{LEE}$ as a function of time since $t_0$, the last moment the detector was at \unit[4]{K}, for each measurement, fitted with a power law model $R_{LEE}(t) = A \cdot (t - t_0)^{-k}$. Dashed curves indicate the best fit power laws; shaded bands represent fit uncertainties. \textbf{Top right:} The fitted power law exponents $k$ for each run; the shaded area indicates the weighted average $\overline{k} = 0.59 \pm 0.06$. \textbf{Bottom right:} Fit amplitudes $A$ from power law fits with fixed exponent $k=\overline{k} = 0.59$, plotted as a function of the cooldown duration from room temperature to the last moment at \unit[4]{K}.}
        \label{fig:rate_vs_time}
\end{figure*}
\begin{figure}
        \centering
        \includegraphics[width=1\linewidth]{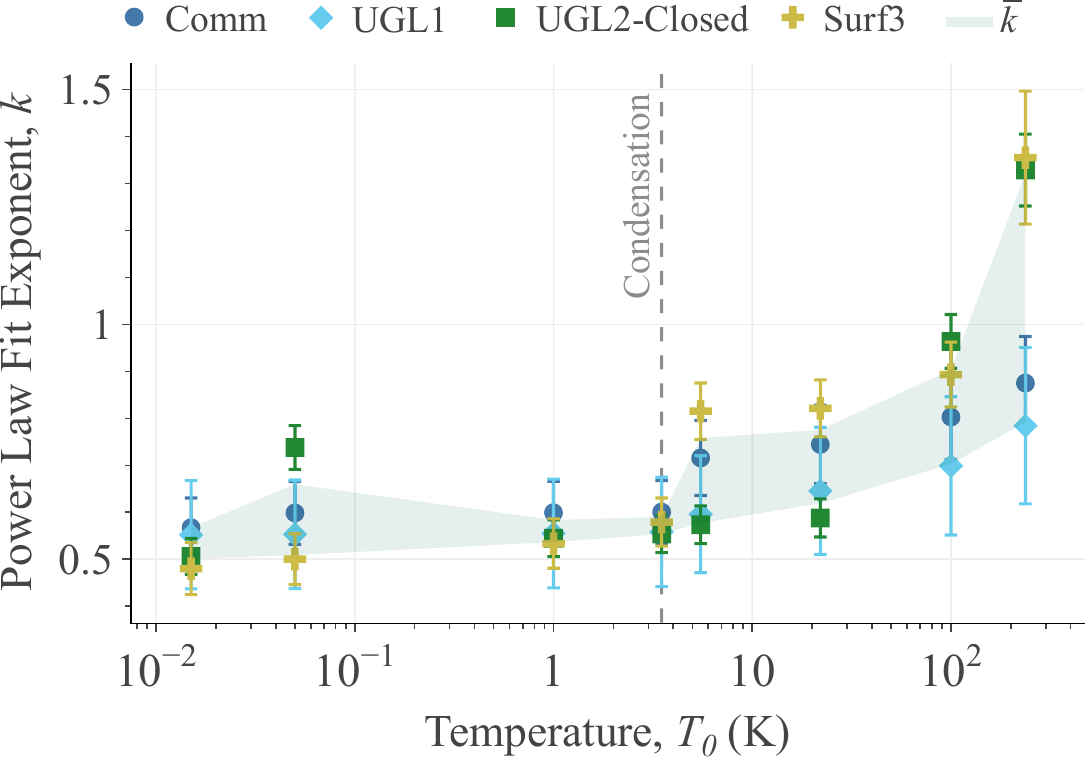}
        \caption{Power law exponent $k$ obtained from fits of the LEE rate evolution with $R(t) = A \cdot (t - t_0)^{-k}$ using different definitions of the reference time $t_0 = t(T=T_0)$, where $T_0$ is the chosen anchor temperature. The shaded band indicates the standard deviation from the weighted average of the resulting $k$ values across runs. The spread in the fitted values is minimal when $t_0$ is defined at the start of condensation (dashed line), supporting the choice of \unit[4]{K} as a consistent reference point across runs.}
        \label{fig:temp_scan}
\end{figure}
        
A gradual decrease in the LEE rate over time was first observed by the EDELWEISS Collaboration~\cite{queguiner_edelweiss_thesis:2018} and has since been reported in multiple studies by CRESST~\cite{CRESST_LEE:2023,CRESST_2TES:2024} and TESSERACT~\cite{TESSERACT_stress:2024,tesseract_mass}. Moreover, it was shown that the LEE rate increases after a warm-up of the detector system to tens of Kelvin~\cite{queguiner_edelweiss_thesis:2018,CRESST_LEE:2023}, highlighting the relevance of temperature changes in the LEE rate. In this section, we examine the time evolution of the shared LEE events observed with the NUCLEUS double TES detector, taking into account the temperature profile during the cryostat cooldown, like the one shown in Figure~\ref{fig:cooldown_curve}, as a possible factor affecting the LEE rate. 

The left panel of Figure~\ref{fig:rate_vs_time} shows the evolution over time of the LEE event rate, as defined in Eq.~\ref{eq:rlee}. The uncertainties on the LEE rate include both a systematic component derived from the difference between the two independent analyses and statistical uncertainties from each analysis. In all runs with a data-taking period spanning more than 2 days (Surf3, UGL1, Comm, UGL2-Closed), a clear decrease in the event rate is observed.

In order to perform a search for a critical temperature affecting the LEE, we model the rate evolution with the following power law function:
\begin{equation}\label{eq:powerlaw}
    R_{LEE}(t) = A \cdot (t - t_0)^{-k},
\end{equation}
where $t_0$ is the moment in time when the detector reaches the temperature $T_0$, $A$ is a normalization constant, $k$ is the exponent indicating the LEE time decay, and $t$ is the time at which the data point is measured. The sensitivity to a precise instant $t_0$ makes the power law model a useful tool to characterize the LEE. 

The choice of $T_0$, and thus $t_0$, was made with a data-driven approach. While the ideal time reference would be physically motivated by the underlying origin of the LEE, this approach could, in turn, help identify the relevant physical processes. Following the time evolution of the temperature recorded by the thermometer mounted on the box enclosing the detector, shown in Figure~\ref{fig:cooldown_curve}, we tested different definitions of $t_0(T_0)$ and extracted the corresponding values of $k$ by fitting Eq.~\ref{eq:powerlaw} to the LEE rate in different runs. As visible from Figure~\ref{fig:temp_scan}, the spread in $k$ is largest when $t_0$ is defined as the start of the pulse tube, i.e., the beginning of the cooldown from room temperature. In this case, the fitted exponents vary significantly across runs, suggesting that the LEE rate would follow different power laws for different datasets. Given that the same detector was used, such a scenario appears unlikely. While the agreement is not optimal for higher anchor temperatures, the fitted exponents become significantly more consistent when $T_0$ is around \unit[4]{K} or lower. In the NUCLEUS dry dilution refrigerator, this temperature indicates the start of the condensation of the helium mixture, a process that brings the system to the base temperature of \unit[10]{mK}. Since the extracted $k$ parameters are most consistent when $t_0 = t_0(\unit[4]{K})$, we chose this value as the start of the timescale. It is worth noting that the time required to reach \unit[10]{mK} from \unit[4]{K} is, in every dataset, small compared to the start of the data-taking, making any temperature choice below \unit[4]{K} fairly equivalent as visible from Figure~\ref{fig:temp_scan}.

The results of fitting Eq.~\ref{eq:powerlaw} with $t_0 = t_0(\unit[4]{K})$ are shown as dashed lines in the left panel of Figure~\ref{fig:rate_vs_time}. The fit was performed using orthogonal distance regression, which accounts for uncertainties in both variables (rate and time) by minimizing orthogonal residuals, corresponding to a generalized $\chi^2$. The top-right panel displays the extracted power law exponent $k$ for each dataset as a function of the cooldown time from room temperature to \unit[4]{K}. The measurements yield consistent values, with a weighted average of $\overline{k} = 0.59 \pm 0.06$, when the time difference $t-t_0$ is expressed in days. In the bottom right panel, instead, we show the behavior of the normalization constant $A$, obtained by performing a new fit using a fixed exponent $k = \overline{k} = 0.59$, with respect to the time it took to reach the \unit[4]{K} temperature from the beginning of the cooldown.

Moreover, during the NUCLEUS commissioning run a \cawo single TES detector was present along with the \alo detector~\cite{LBR_paper}.  By performing the same time dependence analysis described before, the measured time decay exponent for \cawo is $k_{\text{CaWO}_4}=0.73 \pm 0.08$. This value is compatible within 1.5~$\sigma$ with the average measured $\overline{k}=0.58\pm 0.06$ for \alo. These observations further support the interpretation that the LEE is primarily driven by cooldown-related effects that are identical for the two detectors.

The compatibility of the extracted values of $k$ can be interpreted as follows: the LEE rate is initially set to a value proportional to $A$ when the detector reaches \unit[4]{K}, and decreases from that point on with the same time scaling law. The extracted amplitudes $A$, which represent the LEE rates one day after \unit[4]{K}, show a strong decreasing trend with increasing duration of the cooldown from room temperature to the condensation start. In other words, the NUCLEUS data obtained with the same detector but under different cooldown conditions suggest that slower cooldowns lead to lower initial LEE rates, by up to an order of magnitude, while the subsequent time evolution is comparable across the runs. This finding indicates that the initial LEE rate can potentially be further reduced by adjusting the cooldown parameters, particularly if the critical phase of the cooldown is identified.

Several different processes freeze-out at temperatures of O(\unit[1]{K}) that could possibly affect the LEE rate. For example, thermal contractions of the copper surrounding the detector stops at O(\unit[10]{K})~\cite{thermalcontraction}, possibly causing a freeze-out of the stress induced on the detector which is known to affect the level of the LEE~\cite{TESSERACT_stress:2024}. Another phenomenon that happens at around \unit[1]{K} is the condensation of helium; the amount of helium remaining in the vacuum of the cryostat can change as long as it is present in gas form, but it is impossible to remove after condensation. The presence of liquid helium on top of the detector surface could impact the LEE rate due to the extremely low viscosity of the fluid and its tendency of spreading on surfaces and penetrating inside crevices. Moreover, at around \unit[1]{K} all the aluminum near the detector, such as the wiring and the phonon collecting pads of the TESs, becomes superconductive. The conditions in which such a sharp transition happens, i.e. magnetic flux or the spatial homogeneity of the process, could impact the rate recorded by the detector.

The LEE rate has often been modeled using an exponential decay~\cite{CRESST_LEE:2023,CRESST_2TES:2024,TESSERACT_stress:2024}; we also tried this approach, but with the data presented in this work the extracted decay constants vary significantly between cooldowns, spanning a broad range from 3 to 30 days. This is a further motivation to describe the LEE time decay with the power law model of Eq.~\ref{eq:powerlaw}, since it allows a more consistent description across datasets. Despite the wide range in decay constants, the quality of the performed exponential fits, evaluated using the reduced $\chi^2$ method, remains in line with the power law one, meaning that both models are equally statistically valid.

\section{Conclusions and outlook} 

In this work we have presented a systematic study of the LEE observed with the NUCLEUS \alo double TES detector operated under various experimental conditions. We focused on LEE events in the energy range between 100 and \unit[300]{eV}, and observed in coincidence between the two sensors, as they mimic the expected \cevns signal originating from energy deposition in the bulk of the crystal. The results allow us to exclude several potential origins of the excess in the low energy counting rate. In particular, the observed LEE rates are incompatible with noise-induced false triggers and cannot be attributed to muon interactions, as demonstrated through the first dedicated coincidence analyses of cryogenic detector data with a muon veto in a high muon flux environment. Furthermore, a dedicated test with opening the external shield within the same run, as well as comparisons between runs before and after cleaning the setup, show that the LEE rate does not correlate with the particle background level. 

A key finding of this study is the dependence of the initial LEE rate on the cooldown dynamics. Runs with slower cooldowns consistently exhibited lower initial LEE rates, while the subsequent time evolution of the LEE rate followed a universal power law behavior,  $R_{LEE}(t) = A \cdot (t-t_0)^{-k}$, with $(t-t_0)$ defined as the time, in days, since the start of condensation, corresponding to the temperature of about \unit[4]{K}. The exponent $k = 0.59 \pm 0.06$ is consistent across all runs, while the normalization $A$, corresponding to the LEE rate after one day at \unit[4]{K}, decreases with increasing cooldown duration (time required to reach \unit[4]{K} from room temperature).  

These results provide strong evidence that the LEE is driven neither by particle background nor by noise, and point toward cooldown related processes as a key factor in shaping the initial rate. Further systematic investigations of the impact of cooldown dynamics will help to identify the most critical temperature range for the processes triggering the LEE and, in turn, possibly point to the underlying physical mechanisms. Controlling the cooldown speed within this range may provide an effective experimental strategy to reduce the LEE rate.

Further dependencies of the LEE rate with respect to the detector history have been searched and are presented in Appendix~\ref{app:remount}. The results yield no clear dependence between the LEE rate and the following: time since the last crystal remounting, the number of thermal cycles, or the total time spent at cryogenic temperature. This indicates that remounting procedures do not play a significant role in the evolution of the LEE rate.

It is worth mentioning that the negative trigger rate, shown in Figure~\ref{fig:energy_spectra}, affecting the region-of-interest of NUCLEUS is expected to be drastically lowered in future data-taking campaigns. This expectation stems from the fact that all the negative triggers shown happen after an increase in the noise affecting the data, as presented in \cite{LBR_paper}. Moreover, both sensors show a correlated noise at low frequency, meaning that fluctuations are not independent between the two TESs. A systematic study of the noise correlation between TESs sensing the same crystal will be the subject of an upcoming publication, along with new analysis methods that allow for lowering its negative impact. 

Finally, although discrimination between \textit{singles} and shared events is possible with the multi-sensor readout, this separation is limited and becomes increasingly difficult at the lowest energies, where the \textit{singles} LEE contribution dominates. Future dedicated studies of the single-channel LEE events are therefore crucial, as understanding their origin may offer pathways to reliably reduce the LEE at the lowest energies.

\backmatter

\bmhead{Acknowledgements}
This work has been financed by the CEA, the INFN, the {\"O}AW and partially supported by the TU Munich and the MPI f{\"u}r Physik. NUCLEUS members acknowledge additional funding by the DFG through the SFB1258 and the Excellence Cluster ORIGINS, by the European Commission through the ERC-StG2018-804228 "NUCLEUS", by the P2IO LabEx (ANR-10-LABX-0038) in the framework "Investissements d’Avenir" (ANR-11-IDEX-0003-01) managed by the Agence Nationale de la Recherche (ANR), France, by the Austrian Science Fund (FWF) through the "P 34778-N, ELOISE", and by Max-Planck-Institut f{\"u}r Kernphysik (MPIK), Germany.

\bibliography{sample}

\newpage

\begin{appendices}
\bmhead{Appendix A: Excluded Correlations with Detector History}\label{app:remount}

\begin{figure*}[t]
        \centering
        \includegraphics[width=1\linewidth]{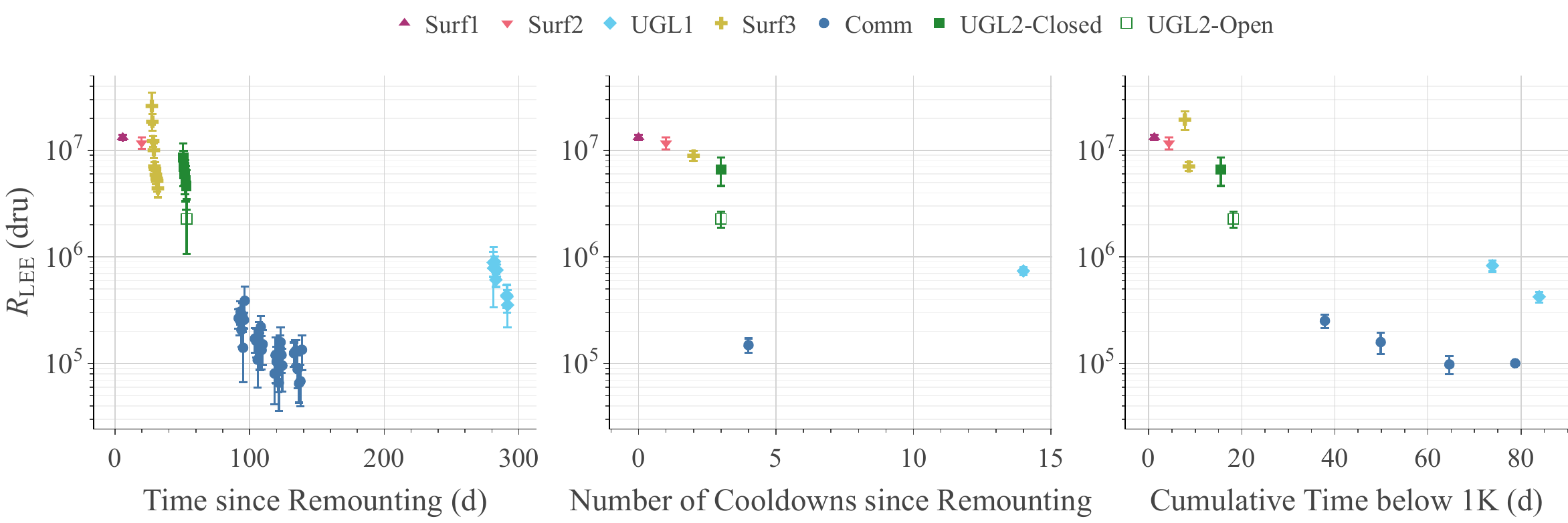}
         \caption{Additional attempted studies showing broken or inconsistent correlations, particularly evident when comparing the commissioning run (\textit{Comm}) and \textit{UGL1} datasets. \textbf{Left}: LEE rate as a function of the total time elapsed since the last detector mounting. \textbf{Center}: LEE rate versus the number of thermal cycles the detector underwent since its last mounting in the holder. Average LEE rates for each dataset are shown. \textbf{Right}: Evolutions of the LEE while the detector remains at cryogenic temperature below \unit[1]{K} are investigated and excluded. For clarity, long measurements are divided into several periods, and the average LEE rate in each period is shown.}
        \label{fig:LEE_discarded}
\end{figure*}

Aside from the correlations of the LEE rate with cooldown parameters presented in the main text, other variables have been investigated and resulted in unconvincing correlations. The results of these studies are shown in Figure~\ref{fig:LEE_discarded}, with the aim of reducing parameter degeneracies in the characterization of the LEE.

In Figure~\ref{fig:LEE_discarded} (Left), we present the LEE rate as a function of the total time elapsed since the detector was last mounted in its holder. The observation of an approximately three times higher LEE rate in the UGL1 run (280 days since detector remounting) compared to the Comm run (90 days since remounting) is inconsistent with a correlation between the LEE rate and the total time elapsed since the detector was last mounted in its holder. This indicates that keeping the detector mounted at room temperature does not lead to a reduction of the LEE, contrary to one of the hypotheses previously proposed by the community.

In Figure~\ref{fig:LEE_discarded} (central), we show the LEE rate as a function of the number of cooldowns the detector underwent since its last remounting. The motivation to look at the correlation between LEE rate and the number of cooldowns is given by the hypothesis that repeated thermal cycles could cause stress annealing in the crystal, either from internal deformations or from mounting-induced stress. However, the comparison between the UGL1 and Comm runs shows no correlation, so we see no indication that repeated thermal cycling would reduce the LEE rate.

A correlation between LEE rate and the cumulative time the detector spent below \unit[1]{K} was also investigated, motivated by the hypothesis that the LEE decays only at low temperatures, which may be inferred from the studies in \cite{CRESST_LEE:2023,phdthesis_Underwood}. Figure~\ref{fig:LEE_discarded} (Right) shows the behavior of the LEE with respect to the cumulative time below \unit[1]{K}. As in the previous cases, the comparison of LEE rates from the UGL1 and Comm runs does not support this hypothesis, indicating that the cooldown speed remains the most promising handle on the LEE rate identified in this study.

\bmhead{Appendix B: Comparison with LEEs Measurements in Other Cryogenic Experiments}\label{app:remount}

In the commissioning run of the NUCLEUS experiment (\textit{Comm})~\cite{LBR_paper}, both the \alo and \cawo NUCLEUS detectors recorded the lowest LEE observed so far by the collaboration. Moreover, the spectra recorded in the two detectors appear remarkably similar, despite the different target materials. In this appendix, we place these results in the broader context of LEE measurements reported by other experiments employing cryogenic calorimetry.
\begin{figure*}
        \centering
        \includegraphics[width=1.05\linewidth]{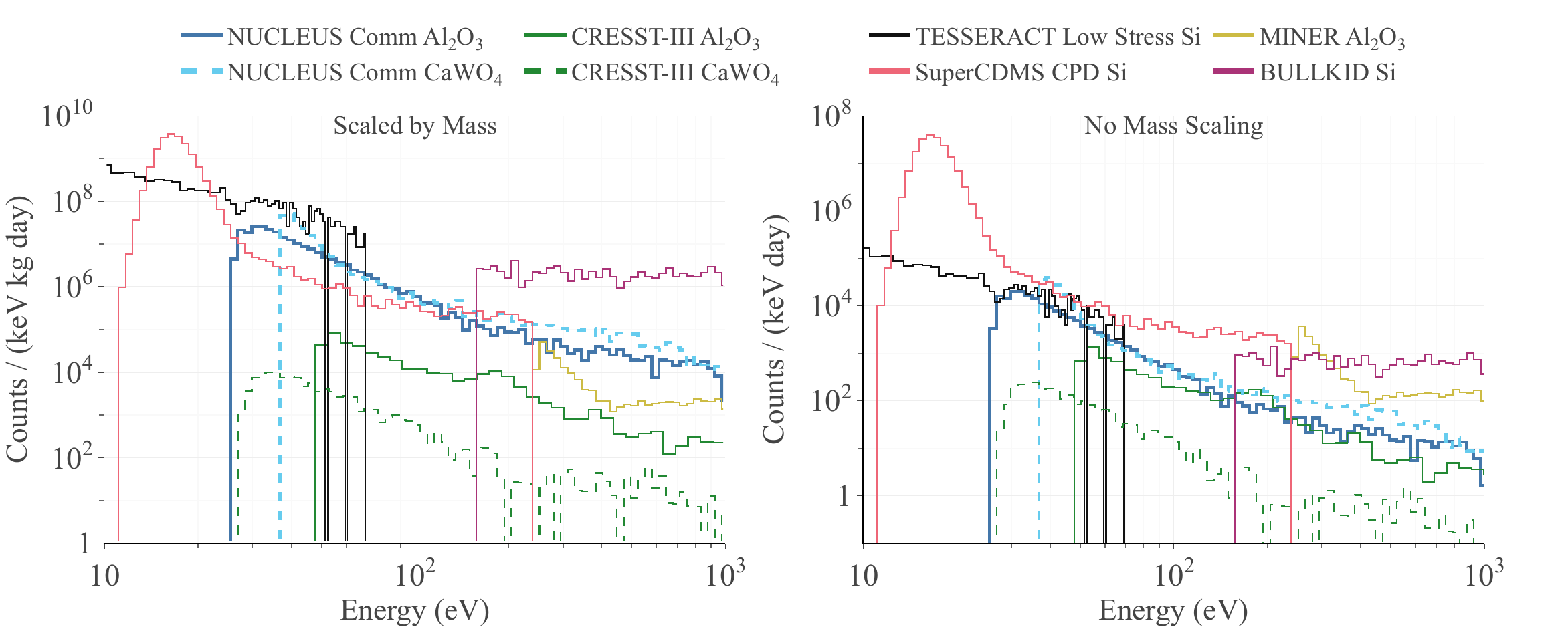}
        \caption{Comparison of low energy spectra measured by experiments employing cryogenic calorimetry. Spectra from the \alo and \cawo CRESST detectors~\cite{CRESST_LEE:2023}, operated deep underground at LNGS, and from the surface-based experiments SuperCDMS-CPD~\cite{SCDMS_CPD}, TESSERACT~\cite{TESSERACT_stress:2024}, MINER~\cite{miner}, and BULLKID~\cite{bullkid2024}, as well as from the NUCLEUS commissioning run \alo (this work) and \cawo~\cite{LBR_paper} detectors, are shown with (left) and without (right) normalization to the crystal mass. The NUCLEUS spectra are presented without applying the anticoincidence cut with the muon-veto signals.
        }
        \label{fig:LEE_FULL}
\end{figure*}
Figure~\ref{fig:LEE_FULL} (Left) shows the low energy spectra expressed in the standard units used in CE$\nu$NS and dark matter searches, with the event rates normalized by the detector mass. For comparison, we also include representative measurements from other cryogenic experiments: CRESST~\cite{CRESST_LEE:2023}, which pioneered the TES-based technology later adopted by NUCLEUS; TESSERACT~\cite{TESSERACT_stress:2024}, SuperCDMS-CPD~\cite{SCDMS_CPD}, and MINER~\cite{miner}, which use quasiparticle-assisted electrothermal feedback transition-edge sensors (QETs)~\cite{QET}; and BULLKID~\cite{bullkid2024}, based on Kinetic inductance detectors (KIDs) phonon-sensing technology~\cite{Day2003}.

A long-standing question within the LEE community is whether the excess rate scales with the mass of the crystal substrate. Different experiments have reported partially conflicting indications on this point, notably CRESST~\cite{CRESST_LEE:2023} and TESSERACT~\cite{tesseract_mass}. To illustrate this issue, Figure~\ref{fig:LEE_FULL} (Right) shows the same low energy spectra but without applying any mass normalization. In this representation, the measured LEE rates appear more similar across experiments than in the mass-scaled comparison, suggesting that a simple scaling with detector mass may not fully capture the underlying behavior of the excess. 

Detectors operated in deep underground laboratories have generally exhibited lower LEE rates than surface-based experiments~\cite{EXCESS_review:2022}. The reason can be tracked down to the fact that underground experiments usually operate with background shields at cryogenic temperatures. This additional payload induces a slower cooldown procedure which then causes a delay of the start of the setup characterization and the data taking; both these factors lower the LEE rate as seen in Section~\ref{sec:rate_vs_time}. This, in fact, is the same type of degeneracy described in Section~\ref{sec:particle_bkg}. In this context, it is noteworthy that the LEE rate measured during the NUCLEUS commissioning run, which featured a relatively long cooldown (see Figure~\ref{fig:cooldown_curve}) despite operating with a background of around $\unit[10^3]{counts/(keV\,kg\,d)}$, is comparable to that observed in CRESST \alo detectors operated at LNGS with a background of O($\unit[10]{counts/(keV\,kg\,d)}$), when the spectra are compared without mass normalization.

On the other hand, if effects related to the nuclear mass or composition of the target crystal play a significant role, the discrepancy between the CRESST-III \cawo module and the NUCLEUS \cawo detector visible in both representations of Figure~\ref{fig:LEE_FULL} remains unexplained. This contrast is particularly striking given the good agreement observed between the \alo detectors of NUCLEUS and CRESST. Further dedicated studies will be required to disentangle the relative contributions of detector material, mounting, and cooldown history to the observed LEE rates.

It is worth mentioning that the above considerations, based on the panels of Figure~\ref{fig:LEE_FULL}, do not take into account any cooldown duration correction, since this information is not commonly available. Moreover, plots like the one in Figure~\ref{fig:LEE_FULL} have been the community standard used to compare difference experiments (e.g. in \cite{EXCESS_review:2022,bullkid2024,EXCESS_review:2025}).

\end{appendices}
\end{document}